\documentclass[11pt]{article}
\usepackage{amsmath}
\usepackage{amssymb}
\usepackage{amsthm,amsxtra}
\usepackage{epsf}
\usepackage{eepic}
\newlength{\mytopmargin}
\newlength{\myleftmargin}
\renewcommand{\theequation}{\thesection.\arabic{equation}}
\newtheorem{lemma}{Lemma}
\newtheorem{prop}[lemma]{Proposition}
\newtheorem{thm}{Theorem}
\newtheorem{cor}[thm]{Corollary}

\begin{document}

\title{\Large\bf Determinantal Correlations for Classical Projection Processes}
\author{Peter J. Forrester$^{\dagger}$ and Taro Nagao${}^*$}
\date{}
\maketitle

\begin{center}
\it
 $^{\dagger}$Department of Mathematics and Statistics, University of Melbourne, \\
Victoria 3010, Australia
\end{center}
\begin{center}
${}^*$
\it
Graduate School of Mathematics, Nagoya University, \\
Chikusa-ku, Nagoya 464-8602, Japan
\end{center} 

\bigskip
\begin{center}
\bf Abstract 
\end{center}
\par
\bigskip
\noindent 
Recent applications in queuing theory and
statistical mechanics have isolated the process formed by the eigenvalues
of successive minors of the GUE. Analogous eigenvalue processes, formed
in general from the eigenvalues of nested sequences of matrices resulting
from random corank 1 projections of classical random matrix ensembles, are 
identified for the LUE and JUE. The correlations for all these processes
can be computed in a unified way. The resulting expressions can then be 
analyzed in various scaling limits. At the soft edge, with the rank of the
minors differing by an amount proportional to $N^{2/3}$, the scaled
correlations coincide with those known from the soft edge scaling of the 
Dyson Brownian motion model.

\newpage
\section{Introduction}
Since the pioneering days of Wigner, Gaudin, Mehta and Dyson (see Porter \cite{Po65} for a collection
of papers from this period), random matrix theory has shown itself to be perhaps the richest 
source of exact solutions for correlation and distribution functions of all statistical
mechanical models. Many of the discoveries of this type have their motivation in new applications
of random matrix theory, unknown in the pioneering days. A case in point is the recent work of
Johansson and Nordenstam \cite{JN06,No07}, who compute the exact form of the correlation functions
for the coupled eigenvalue sequences obtained from the principal minors of Gaussian unitary
ensemble (GUE) matrices.

The motivation for studying this GUE minor process begins with a work of Baryshnikov \cite{Ba01}.
Some years earlier Glynn and Whitt \cite{GW91} had studied the problem of computing the distribution
of exit times from a queueing system, in the limit the number of queues tends to infinity but the
number of jobs remains finite. It was proved that for general i.i.d.~service times the scaled
distributions $D_k$ of the exit time of the $k$th customer from the final queue could be written as
$$
D_k = \sup_X \sum_{i=0}^{k-1} \Big ( B_i(t_{i+1}) - B_i(t_i) \Big ).
$$
Here each $B_i$ denotes an independent standard Brownian motion, while the condition $X$ is that
$$
0 = t_0 < t_1 < \cdots < t_k = 1.
$$
By studying the particular case of exponential waiting times Baryshnikov was able to show that
$\{D_k\}_{k=1,2,\dots}$ could alternatively be specified as the joint distribution of $\{\mu_k\}$,
where $\mu_k$ is the largest eigenvalue of the $k$th principal minor of an infinite GUE matrix
$X$
with probability density function (PDF) proportional to $\exp(-X^2/2)$.
Johansson and Nordenstam \cite{JN06} give other occurrences in statistical mechanics of essentially
the same process
relating to the eigenvalues of minors of GUE matrices (referred to as the GUE minor process). 
These are in the specification of certain point processes
relating to the boundary region (neighbourhood of the frozen zone) of random domino tilings of the
Aztec diamond \cite{Jo05a}, and to random lozenge-tilings of a hexagon \cite{Jo05b}. In the
equivalent language of stepped surfaces, Okounkov and Reshetikhin \cite{OR06} make similar
observations.
The recent work of Borodin, Ferrari and Sasamoto \cite{BFS07a} encounters this process in the context
of studying the dynamics of the asymmetric exclusion process.

The broader setting of the measures encountered in the studies of the above statistical mechanical models
relates to the Robinson-Schensted-Knuth (RSK) correspondence (see Section \ref{s2.1}) below.
Consideration of the structures inherent therein \cite{BR01a,FR02b} identifies natural extensions of
the GUE minor process. One such class of extensions replaces the Gaussian weight in the latter
by the Laguerre or Jacobi weights, which can all be realized by a sequence of projections onto random
complex hyperplanes, giving rise to so called classical projection processes (the Gaussian,
Laguerre and Jacobi weights are all classical from the viewpoint of the theory of orthogonal polynomials).
Our main point in the present paper
is that the correlations for the classical projection processes can be computed exactly in
a unified way. The essential ingredient here is the Rodrigues formula for classical orthogonal
polynomials. Moreover, known asymptotic formulas for the latter allow for the evaluation of
scaling limits of the correlations.

We begin in Section 2 by recalling how the RSK correspondence relates to a certain statistical mechanical
model of last passage times. In Section 3 the occurrence of special cases of the joint 
PDF in some random matrix setting is noted. The correlations for the classical
projection process are calculated in Section 4, and their scaling limits analyzed in
Section 5.

\section{A joint probability density associated with RSK}
\setcounter{equation}{0}

\subsection{The case of general parameters}\label{s2.1}
The Robinson-Schensted-Knuth (RSK) correspondence gives a bijection between $n_1 \times
n_2$ non-negative integer matrices $[x_{i,j}]$ (rows countered from the bottom) with entry
$(ij)$ weighted $(a_i b_j)^{x_{i,j}}$, and pairs of weighted semi-standard tableaux (weights
$\{a_i\}$, $\{b_j\}$) of shape $\mu=(\mu_1,\dots,\mu_n)$). Results from \cite{Jo99a,BR01a} associate
a probabilistic model to the RSK correspondence. Identify with each lattice site $(ij)$ a
random non-negative integer variable $x_{i,j}$ chosen from the geometric distribution with
parameter $a_i b_j$, so that
\begin{equation}\label{Pab}
{\rm Pr}(x_{i,j} = k) = (1 - a_i b_j) (a_i b_j)^k.
\end{equation}
For given non-negative parameters $\{a_i\}$, $\{b_j\}$, and given $n_1, n_2 \in \mathbb Z^+$, a
probabilistic quantity of interest is the sequence of last passage times
\begin{equation}\label{Ll}
L^{(l)}(n_1,n_2) = \max \sum_{({\rm rd}^*)^l} x_{i,j}, \qquad
l=1,\dots,\max (n_1,n_2) .
\end{equation}
Here (rd${}^*)^l$ denotes the set of $l$ disjoint (no common lattice points) rd${}^*$ lattice paths,
which in turn are defined as either a single point, or points connected by segments formed out of
arbitrary positive integer multiples of steps to the right and steps up in the rectangle
$1 \le i \le n_1$, $1 \le j \le n_2$.

A crucial feature of the RSK correspondence is that the length of the first row of the semi-standard
tableaux pair is equal to (\ref{Ll}) in the case $l=1$, and thus $\mu_1 = L^{(1)}(n_1,n_2)$.
More generally, all row lengths are determined by (\ref{Ll}) according to \cite{Gr74} 
\begin{equation}\label{Ll1}
\mu_l = L^{(l)}(n_1,n_2) - L^{(l-1)}(n_1,n_2)
\end{equation}
with $L^{(0)}(n_1,n_2) := 0$. Thus the distribution of the last passage times is fully determined
by the distribution of $\{\mu_l\}$.

A well known fact relating to the RSK  correspondence is that the probability an $n_1 \times n_2$
non-negative matrix with elements chosen according to (\ref{Pab}) corresponds to a pair of
semi-standard tableaux with shape $\mu$, one of content $n_1$, the other of content $n_2$, is given
by \cite{Kn70}
\begin{equation}\label{2.3}
P(\mu) = \prod_{i=1}^{n_1} \prod_{j=1}^{n_2} (1 - a_i b_j)
s_\mu(a_1,\dots,a_{n_1}) s_\mu(b_1,\dots,b_{n_2}),
\end{equation}
where $s_\mu$ denotes the Schur polynomial.
A lesser known fact is that for $n_1 > n_2$ the joint probability that an
$n_1 \times (n_2 + 1)$ non-negative integer matrix with elements chosen according
to (\ref{Pab}) corresponds to a pair of semi-standard tableaux with shape $\mu$,
content $n_1$ and $n_2 + 1$, and the bottom left sub-matrix corresponds to a pair
of semi-standard tableaux with shape $\kappa$, content $n_1$ and $n_2$ is \cite{FR02b}
\begin{equation}\label{2.4}
\prod_{i=1}^{n_1} \prod_{j=1}^{n_2+1} (1 - a_i b_j)
s_\mu(a_1,\dots,a_{n_1}) s_\kappa(b_1,\dots,b_{n_2})
b_{n_2+1}^{\sum_{j=1}^{n_2} (\mu_j - \kappa_j) + \mu_{n_2+1}} \chi(\mu>\kappa)
\end{equation}
where, with $\chi(A) = 1$ if $A$ is true, $\chi(A) = 0$ otherwise,
\begin{equation}\label{2.11}
\chi(\mu>\kappa) := \chi(
\mu_1 \ge \kappa_1 \ge \mu_2 \ge \kappa_2 \ge \cdots
\ge \mu_{n_1} \ge \kappa_{n_1} \ge 0).
\end{equation}

It follows from (\ref{2.3}) and (\ref{2.4}) that given the pair of semi-standard
tableaux corresponding to an $n_1 \times n_2$ matrix $(n_1 > n_2)$
has shape $\kappa$, the
probability of the pair of semi-standard tableaux corresponding to the
$n_1 \times (n_2+1)$ matrix, obtained by adding an extra row to the existing matrix,
having shape $\mu$ is
\begin{equation}\label{2.8}
P(\mu,\kappa) := \chi(\mu>\kappa) \prod_{i=1}^{n_1} (1 - a_i b_{n_2+1})
{s_\mu(a_1,\dots,a_{n_1}) \over s_\kappa(a_1,\dots,a_{n_1}) }
b_{n_2 + 1}^{\sum_{j=1}^{n_2} (\mu_j - \kappa_j) + \mu_{n_2 + 1} }.
\end{equation}
Let us now seek the joint probability that with $n_1 \ge n_2 + p$, an $n_1 \times (n_2 + p)$
non-negative integer matrix with elements chosen according to (\ref{Pab}) is such that the
principal $n_1 \times (n_2 + s)$ sub-blocks $(s=0,1,\dots,p)$ correspond to pairs of
semi-standard tableaux with shape $\mu^{(s)}$. This is computed from (\ref{2.3}) and (\ref{2.8})
according to
\begin{eqnarray}\label{E1}
 P(\mu^{(0)}) \prod_{s=0}^{p-1} P(\mu^{(s+1)}, \mu^{(s)}) 
& = &
\prod_{i=1}^{n_1} \prod_{j=1}^{n_2+p} (1 - a_i b_j)
s_{\mu^{(p)}} (a_1,\dots,a_{n_1}) s_{\mu^{(0)}}(b_1,\dots,b_{n_2})  \nonumber \\
&&  \times
\prod_{s=1}^p b_{n_2 + s}^{\sum_{j=1}^{n_2+s-1}(\mu_j^{(s)} - \mu_j^{(s-1)}) +
\mu_{n_2 + s}^{(s)} } \chi(\mu^{(s)}>\mu^{(s-1)}).
\end{eqnarray}

\subsection{Specializing the parameters}
In \cite{FR02b} the joint probability (\ref{2.4}) is specialized to the case of a geometrical
progression of parameters
\begin{eqnarray}\label{2.17}
(a_1,\dots,a_{n_1}) & = & (z,zt,zt^2,\dots,zt^{n_1-1}) \nonumber \\
(b_1,\dots,b_{n_2}) & = & (z,zt,zt^2,\dots,zt^{n_2-1}).
\end{eqnarray}
This is a preliminary step for taking the so called Jacobi limit, in which a joint 
probability of a type known from random matrix theory is obtained. In this subsection the
parameters will be specialized according to (\ref{2.17}) for the more general joint
probability (\ref{E1}).

Now for (\ref{E1}) to be non-zero we require $\ell(\mu^{(p)}) \le n_2 + p$. Under this
circumstance, we deduce from \cite[Eq.~(2.26)]{FR02b} with $\kappa \mapsto \mu^{(p)}$,
$n_2 \mapsto n_1$, $n_1 \mapsto n_2+p$, $r_j \mapsto h_j^{(p)} :=
\mu_j^{(p)} + n_2 + p - j$ that
\begin{eqnarray}\label{E2}
 s_{\mu^{(p)}}(1,t,\dots,t^{n_1-1}) 
& = &
t^{-\sum_{j=1}^{n_1-n_2-p} j(j-1)} t^{-(n_2+p) \sum_{j=1}^{n_1-n_2-p} j}
{t ^{- \sum_{j=1}^{n_1} (j-1) (n_2+p-j)} \over \prod_{l=1}^{n_1-1} (t;t)_l }
\nonumber \\
&&  \times \prod_{i=1}^{n_1 - n_2 - p - 1} (t;t)_i
\prod_{i=1}^{n_2+p} {(t;t)_{h_i^{(p)} + n_1 - n_2 - p} \over (t;t)_{h_i^{(p)}} }
\prod_{i < j}^{n_2+p} (t^{h_j^{(p)}} - t^{h_i^{(p)}} ).
\end{eqnarray}
This makes explicit the first Schur polynomial factor in (\ref{E1}). In regards to
the second Schur polynomial factor,
making use of  \cite[Eq.~(2.18)]{FR02b} with $n \mapsto n_2$, $n^* \mapsto n_2$,
$\lambda \mapsto \mu^{(0)}$, $h_j \mapsto h_j^{(0)} := \mu_j^{(0)} + n_2  - j$
gives
\begin{equation}\label{E3}
s_{\mu^{(0)}}(1,t,\dots,t^{n_2-1}) =
{t ^{- \sum_{j=1}^{n_2} (j-1) (n_2-j)} \over \prod_{l=1}^{n_2-1} (t;t)_l }
\prod_{i < j}^{n_2} (t^{h_j^{(0)}} - t^{h_i^{(0)}} ).
\end{equation}
Use of these results in (\ref{E1}) allows the following results to be deduced.

\begin{prop}\label{p1}
Let $n_1 \ge n_2 + p$. On each site of the $n_1 \times (n_2 + p)$ square lattice specify a
random non-negative integer $x_{i,j}$ according to the specification
\begin{eqnarray*}
{\rm Pr}(x_{i,j} = k) & = & (1 - z^2 t^{i+j-2}) (z^2 t^{i+j-2})^k, \quad j \le n_2 \nonumber \\
{\rm Pr}(x_{i,n_2+s} = k) & = & (1 - \alpha_s z t^{i-1}) (\alpha_s z t^{i-1})^k \quad
(s=1,\dots,p).
\end{eqnarray*}
Introduce the notations $h_i^{(p)} := \mu_i^{(p)} + n_2 + p - i$ and
$$
\tilde{\chi}(h^{(p)}, h^{(p-1)}) :=
\chi(h_1^{(p)} \ge h_1^{(p-1)} > h_2^{(p)} \ge h_2^{(p-1)} > \cdots >
h_{n_2+p-1}^{(p)} \ge h_{n_2+p-1}^{(p-1)} > h_{n_2+p}^{(p)}).
$$
In this setting, the joint probability that the matrix $[x_{i,j}]_{n_1 \times (n_2+p)}$ is such that
the sub-matrices $[x_{i,j}]_{n_1 \times (n_2+s)}$ $(s=0,\dots,p)$ correspond, under RSK,
to pairs of tableaux of shape $\mu^{(s)}$ has the explicit form
\begin{eqnarray}\label{Ps1}
&& K_{n_1,n_2,p}(\{\alpha_s\},z,t) z^{\sum_{j=1}^{n_2+p} h_j^{(p)} + \sum_{j=1}^{n_2} h_j^{(0)}}
\prod_{s=1}^p \alpha_s^{\sum_{j=1}^{n_2+s} h_j^{(s)} - \sum_{j=1}^{n_2+s-1} h_j^{(s-1)} }
\tilde{\chi}(h^{(s)},h^{(s-1)}) \nonumber \\
&& \times \prod_{i=1}^{n_2+p} {(t;t)_{h_i^{(p)} + n_1 - n_2 - p} \over (t;t)_{h_i^{(p)}} }
\prod_{i < j}^{n_2+p} (t^{h_j^{(p)}} - t^{h_i^{(p)}})
\prod_{i < j}^{n_2} (t^{h_j^{(0)}} - t^{h_i^{(0)}})
\end{eqnarray}
with
\begin{eqnarray*}
K_{n_1,n_2,p}(\{\alpha_j\},z,t) & = &
z^{-2 \sum_{j=1}^{n_2} (n_2 + p - j) } z^{- \sum_{s=1}^p (p-s) } \Big (\prod_{s=1}^p \alpha_s^{s-p} \Big )
t^{- \sum_{j=1}^{n_1 - n_2 - p} j(j-1) }
\nonumber \\
&& \times t^{-(n_2+p)  \sum_{j=1}^{n_1 - n_2 - p} j}
{t^{- \sum_{j=1}^{n_2} (j-1)(n_2-j)} \over \prod_{l=1}^{n_2-1} (t;t)_l}
{t^{- \sum_{j=1}^{n_1} (j-1)(n_2+p-j)} \over \prod_{l=1}^{n_1-1} (t;t)_l}  \nonumber \\
&& \times \prod_{l=1}^{n_1 - n_2 - p - 1} (t;t)_l
\prod_{i=1}^{n_1} \prod_{j=1}^{n_2} (1 - z^2 t^{i+j-2})
\prod_{s=1}^p \prod_{i=1}^{n_1} (1 - \alpha_s z t^{i-1}).
\end{eqnarray*}
\end{prop}

\subsection{The Jacobi limit}
The Jacobi limit of the setting of Proposition \ref{p1} corresponds to each site of
the $n_1 \times (n_2 + p)$ square lattice specifying a non-negative continuous
exponential random variable with site dependent variance $j \le n_2$ given by
\begin{eqnarray}\label{Kj}
{\rm Pr}(x_{i,j} \in [y,y+dy]) & = & (i+j-2+2a) e^{-y(i+j-2+2a)} dy, \qquad j \le n_2 
\nonumber \\
{\rm Pr}(x_{i,n_2 + s} \in [y,y+dy]) & = & (i-1+a+a_s) e^{-y(i-1+a+a_s)} dy, \qquad
(s=1,\dots,p).
\end{eqnarray}
This can be obtained from (\ref{Ps1}) by setting
\begin{equation}\label{LJ}
t = e^{-1/L}, \quad
z = e^{-a/L}, \quad
\alpha_s = e^{-a_s/L}, \quad
h_j^{(s)}/L = x_j^{(s)}
\end{equation}
and taking the limit $L \to \infty$. The joint probability (\ref{Ps1}), scaled by multiplying
by $L^{(1+p)(n_2 + p/2)}$, has the following limiting form.

\begin{cor}
The PDF obtained by the limiting procedure (\ref{LJ}) applied to (\ref{Ps1}) is equal to
\begin{eqnarray}\label{KF}
&& \tilde{K}_{n_1,n_2,p}(\{a_j\},a)
e^{-a (\sum_{j=1}^{n_2+p} x_j^{(p)} + \sum_{j=1}^{n_2} x_j^{(0)}) }
\prod_{s=1}^p e^{- a_s ( \sum_{j=1}^{n_2 + s} x_j^{(s)} - \sum_{j=1}^{n_2 + s - 1} x_j^{(s-1)}) }
\chi (x^{(s)}> x^{(s-1)}) \nonumber \\
&& \times
\prod_{i=1}^{n_2+p} (1 - e^{- x_i^{(p)}})^{n_1 - n_2 - p}
\prod_{1 \le i < j \le n_2 + p} (e^{-x_j^{(p)}} - e^{-x_i^{(p)}} )
\prod_{1 \le i < j \le n_2} (e^{-x_j^{(0)}} - e^{-x_i^{(0)}} )
\end{eqnarray}
where
$$
\tilde{K}_{n_1,n_2,p}(\{a_j\},a) = { \prod_{l=1}^{n_1 - n_2 - p - 1} l! \over
( \prod_{l=1}^{n_1 - 1} l! )  (\prod_{l=1}^{n_2-1} l!) }
\prod_{s=1}^p {\Gamma(a_s + a + n_1) \over \Gamma(a_s + a) } 
\prod_{i=1}^{n_1} {\Gamma(2a + i + n_2 - 1) \over \Gamma(2a + i - 1) }
$$
and
\begin{equation}\label{2.16a}
\chi(x^{(s)}>x^{(s-1)}) :=
\chi(x_1^{(s)} > x_1^{(s-1)} > \cdots > x_{n_2+s-1}^{(s)} > x_{n_2+s-1}^{(s-1)} > x_{n_2+s}^{(s)} > 0).
\end{equation}
\end{cor}

Changing variables $e^{-x_j^{(s)}} = y_j^{(s)}$ the PDF (\ref{KF}) reads
\begin{eqnarray}\label{KF1}
&& \tilde{K}_{n_1,n_2,p}(\{a_j\},a)
\prod_{i=1}^{n_2+p} (y_i^{(p)})^{a-1} (1 - y_i^{(p)})^{n_1-n_2-p} \prod_{j=1}^{n_2}
(y_j^{(0)})^{a} \prod_{s=1}^p \Big ( \chi(y^{(s)}<y^{(s-1)}) \nonumber \\
&& \quad \times 
{\prod_{j=1}^{n_2+s} (y_j^{(s)})^{a_s} \over \prod_{j=1}^{n_2+s-1} (y_j^{(s-1)})^{a_s+1} } \Big )
\prod_{1 \le i < j \le n_2 + p}(y_j^{(p)} - y_i^{(p)})
\prod_{1 \le i < j \le n_2} (y_j^{(0)} - y_i^{(0)})
\end{eqnarray}
where
\begin{equation}\label{2.17a}
\chi(y^{(s)} < y^{(s-1)}) :=
\chi( 0 < y_1^{(s)} < y_1^{(s-1)} < \cdots < y_{n_2+s-1}^{(s-1)} < y_{n_2+s}^{(s)} < 1).
\end{equation} 
In the special case $a_s = a-j$ ($s=1,\dots,p$) this simplifies to the functional form
\begin{equation}\label{KW}
{1 \over C} \prod_{l=1}^{n_2+p} w(y_l^{(p)})
\prod_{1 \le i < j \le n_2 + p} (y_j^{(p)} - y_i^{(p)})
\prod_{1 \le i < j \le n_2} (y_j^{(0)} - y_i^{(0)})
\prod_{s=1}^p \chi(y^{(s)}< y^{(s-1)}),
\end{equation}
with $C$ the normalization ($C$ will be used generally below for this purpose and
so its explicit value may vary from equation to equation)
and $w(y) = y^\alpha (1 - y)^\beta$ for certain $\alpha$ and $\beta$. The
latter is the Jacobi weight, which in a functional form involving Vandermonde products is typical of a
PDF arising in random matrix theory. Indeed, this functional form for each of the Gaussian,
Laguerre and Jacobi weights can be obtained as eigenvalue PDFs.

Before turning to such random matrix interpretations, it should be pointed out that in the setting
of Proposition \ref{p1}, choosing each site of the $n_1 \times (n_2 + p)$ square lattice
according to continuous exponential random variables
\begin{equation}\label{2.19a}
{\rm Pr}(x_{i,j} \in [y,y+dy]) =  e^{- y} dy
\end{equation}
leads to (\ref{KW}) with the particular Laguerre weight $w(y) = y^{n_1-(n_2+p)}e^{-y}$. Further, rescaling
the variables $y_j^{(p)} \mapsto n_1(1 + y_j^{(p)} \sqrt{2/n_1})$ therein, and taking $n_1 \to
\infty$ gives (\ref{KW}) back with the Gaussian weight $w(y) = e^{-y^2}$. In the case $p=1$ both
of these facts are explicitly demonstrated in \cite[Props.~4\&5]{FR02b}; the extension to general
$p$ involves straightforward limiting procedures applied to (\ref{KF1}).

\section{Joint eigenvalue PDF for some nested sequences of random matrices}
\setcounter{equation}{0}
\subsection{Gaussian unitary ensemble}
By definition, matrices $M_N$ from the $N \times N$ GUE satisfy the recurrence
\begin{equation}\label{M1}
M_{N+1} = \left [ \begin{array}{cc} M_N & \vec{w} \\
\vec{w}^\dagger & a \end{array} \right ]
\end{equation}
where $a \sim {\rm N}[0,1/\sqrt{2}]$ and each component $w_j$ of $\vec{w}$ has distribution
$w_j \sim {\rm N}[0,1/2] + i {\rm N}[0,1/2]$. Further, with $U_N$ the unitary matrix which diagonalizes
$M_N$, and thus $M_N = U_N D_N U_N^\dagger$, where $D_N$ is the diagonal matrix of the eigenvalues of
$M_N$, one has
\begin{equation}\label{M2}
\left [ \begin{array}{cc} U_N & \vec{0} \\
\vec{0}^T & 1 \end{array} \right ]
\left [ \begin{array}{cc} M_N & \vec{w} \\
\vec{w}^\dagger & a \end{array} \right ]
\left [ \begin{array}{cc} U_N & \vec{0} \\
\vec{0}^T & 1 \end{array} \right ]^\dagger \sim
\left [ \begin{array}{cc} D_N & \vec{w} \\
\vec{w}^\dagger & a \end{array} \right ].
\end{equation}
This bordered form is the key to studying the joint distribution of the eigenvalues of the
sequence of GUE matrices $\{M_j\}_{j=1,2,\dots}$, or equivalently that of the sequence of
principal minors of a single infinite GUE matrix \cite{Ba01,FR02b,Fo07}.

In particular, it follows from (\ref{M2}) that the characteristic polynomials $p_N(\lambda),
p_{N+1}(\lambda)$ for $M_N,M_{N+1}$ are related by
$$
{p_{N+1}(\lambda) \over p_N(\lambda) } = \lambda - a - \sum_{i=1}^N {|w_i|^2 \over \lambda - 
\lambda_i^{(N)} }
$$
where $\{\lambda_i^{(N)}\}$ denotes the eigenvalues of $M_N$ assumed ordered
$$
\lambda_1^{(N)} < \lambda_2^{(N)} < \cdots < \lambda_N^{(N)}.
$$
 With $\{\lambda_i^{(N)}\}$ regarded as
given the PDF for the zeros of this random rational function, and thus the PDF for the distribution
of the eigenvalues $\{\lambda_i^{(N+1)}\}$ of $M_{N+1}$, can be computed to be equal to 
$$
e^{-\sum_{j=1}^{N+1} (\lambda_j^{(N+1)})^2 }
{\prod_{j<k}^{N+1} ( \lambda_j^{(N+1)} - \lambda_k^{(N+1)} ) \over
\prod_{j<k}^{N} ( \lambda_j^{(N)} - \lambda_k^{(N)} ) }
\chi(\lambda^{(N+1)}< \lambda^{(N)})
$$
where $\chi(\lambda^{(N+1)}< \lambda^{(N)})$ is specified as in (\ref{2.17a}) except that the first
and last inequalities are  replaced by $ - \infty < $ and $< \infty$ respectively.
From this result, together with the fact that the eigenvalue PDF for $N \times N$ GUE matrices
is proportional to
$$
\prod_{l=1}^N e^{- (\lambda_l^{(N)})^2} \prod_{1 \le j < k \le N}
(\lambda_k^{(N)} - \lambda_j^{(N)} )^2,
$$
it follows that the joint eigenvalue PDF for the sequence of GUE matrices $\{M_n\}_{n=N,\dots,N+p}$
is given by (\ref{KW}) with $n_2=N$ and $w(y) = e^{-y^2}$, and the definition  (\ref{2.16a}) of
$\chi(y^{(s)}<y^{(s-1)})$ modified appropriately.

\subsection{Laguerre unitary ensemble}\label{s3.2}
Matrices $A_{(n)}$ from the $N \times N$ LUE with parameter $a=n-N$ are constructed from $n \times N$
rectangular complex Gaussian matrices $X_{(n)}$ with entries N$[0,1/\sqrt{2}]+ i {\rm N}[0,1/\sqrt{2}]$
according to $A_{(n)} = X_{(n)}^\dagger X_{(n)}$. 
Such matrices satisfy the recurrence
\begin{equation}\label{A1}
A_{(n+1)} = A_{(n)} + \vec{x} \vec{x}^\dagger, \qquad A_{(0)} = [0]_{N \times N}
\end{equation}
where $\vec{x}$ is an $N \times 1$ column vector of complex Gaussians.
Our interest is in the case $n \le N$ of $A_{(n)}$, for which there are $N - n$ zero eigenvalues.
Then, after making use too of the invariance of $\vec{x} \vec{x}^\dagger$ by a unitary similarity
transformation, the recursion (\ref{A1}) can be written in the equivalent form
\begin{equation}\label{A2}
 A_{(n+1)} = {\rm diag}(a_1,\dots,a_n, \underbrace{0,\dots,0}_{N-n}) 
+ \vec{x} \vec{x}^\dagger
\end{equation} 
where $\{a_i\}_{i=1,\dots,n}$ are the non-zero eigenvalues of $A_{(n)}$, assumed ordered so
that $0 < a_1 < \cdots < a_n$.
{}From  (\ref{A2}) it follows that the corresponding characteristic polynomials
are such that
\begin{equation}\label{A3}
{\det(\lambda 1_N - A_{(n+1)}) \over \det(\lambda 1_N - A_{(n)}) } =
1 - \sum_{j=1}^n {|x_j|^2 \over \lambda - a_j} -
{\sum_{j=n+1}^N |x_j|^2 \over \lambda}.
\end{equation}

The conditional PDF for the zeros of this random rational function, and thus the
conditional PDF for the eigenvalues of $ A_{(n+1)}$, can be computed exactly as
\cite{FR02b,Fo07}
$$
\prod_{i=1}^{n+1} \lambda_i^{N-(n+1)} \prod_{j=1}^n {1 \over a_j^{N-n} }
e^{-\sum_{j=1}^{n+1} \lambda_j + \sum_{j=1}^n a_j}
{\prod_{i < j}^{n+1} (\lambda_i - \lambda_j) \over
\prod_{i < j}^{n} (a_i - a_j) } \chi (\lambda< a)
$$
Recalling that the non-zero eigenvalue PDF for the matrices $A_{(n)}$ is proportional to
$$
\prod_{j=1}^n a_j^{N-n} e^{-a_j} \prod_{i<k}^n (a_i - a_k)^2
$$
it follows that the joint eigenvalue PDF for the sequence of LUE matrices $\{A_{(r)}\}_{r=n,\dots,n+p}$
with $n+p \le N$ is given by (\ref{KW}) with $n_2 = n$, $w(y) = y^{N-(n+p)} e^{-y}$.

\subsection{Corank 1 random projections}
Let $A_n$ be an $n \times n$ matrix with eigenvalues $a_1 < a_2 < \cdots < a_n$, and let
$\vec{x}$ be an $n \times 1$ random complex Gaussian normalized column vector. The matrix
\begin{equation}\label{MA}
M_n := \Pi_n A_n \Pi_n, \qquad \Pi_n := 1_n - \vec{x} \vec{x}^\dagger
\end{equation}
then represents a corank 1 random projection of $A_n$. We know from \cite{FR02b} that in general
$M_n$ has a single zero eigenvalue, while the non-zero eigenvalues $\{\lambda_j\}_{j=1,\dots,n-1}$ have
the conditional PDF
\begin{equation}\label{CA}
(n-1)! {\prod_{i < j}^{n-1} (\lambda_j - \lambda_k) \over \prod_{i < j}^n (a_j - a_k) }
\chi(a<\lambda).
\end{equation}

Introduce a nested sequence of matrices $\{A_i\}_{i=1,\dots,n}$ by setting $A_{n-1}$ equal to the
diagonal matrix formed from the non-zero eigenvalues of $M_n$, computing $M_{n-1}$ according to
(\ref{MA}), setting $A_{n-2}$ equal to the diagonal matrix formed from the
the non-zero eigenvalues of $M_{n-1}$, and repeating. With the eigenvalues of $A_n$ having PDF
proportional to
\begin{equation}\label{CA1}
\prod_{l=1}^n w(a_l) \prod_{1 \le j < k \le n} (a_k - a_j)^2
\end{equation}
it follows immediately from (\ref{CA}) that the joint PDF for $\{A_i\}_{i=n-p,\dots,n}$ is given by
(\ref{KW}) with $n_2 + p = n$.

We remark (see e.g.~\cite{Fo02})
that the Laguerre and Jacobi cases can be realized by the matrix structure $A_n = X_n^\dagger X_n$
for certain matrices (recall the discussion of the previous subsection for the Laguerre case).
It follows that the above construction is then equivalent to applying a sequence of corank 1 projections
directly to the matrix $X_n$.

\section{Correlation for the classical projection process}
\setcounter{equation}{0}
\subsection{Approach via a general formula}
In studying correlations associated with (\ref{KW}), the most general case occurs when
$n_2=1$. After then changing notation by setting $p=N-1$, $y_l^{(p)} \mapsto x_{p+2-l}^{(p+1)}$,
(\ref{KW}) reads
\begin{equation}\label{EM}
{1 \over C} \prod_{l=1}^{N} w(x_l^{(N)}) \prod_{1 \le j < k \le N} (x_j^{(N)} - x_k^{(N)})
\prod_{s=1}^{N-1} \chi(x^{(s+1)}>x^{(s)}).
\end{equation}
Here $\chi(x^{(s+1)} > x^{(s)})$ is defined as
$$
\chi(x^{(s+1)} > x^{(s)})= 
\chi(x^{(s+1)}_1 > x^{(s)}_1 > \cdots 
> x^{(s+1)}_s > x^{(s)}_s > x^{(s+1)}_{s+1})
$$
(cf.~(\ref{2.16a}))
and $w(x)$ involves a factor $\chi_{0 < x < 1}$.
Hereafter we consider more general cases in which  
$w(x)$ does not always involve such a factor
(see (\ref{3.2}) below).
Of course the product of differences can be written in terms of the Vandermonde determinant, giving
$$
\prod_{l=1}^{N-1} w(x_l^{(N)}) \prod_{1 \le j < k \le N} (x_j^{(N)} - x_k^{(N)})
\propto  \det [ w(x_k^{(N)}) p_{N-j}(x_k^{(N)})]_{j,k=1,\dots,N},
$$
with $\{p_j(x)\}_{j=0,\dots,N-1}$ a set of arbitrary polynomials, $p_j(x)$ of degree $j$.
Furthermore, we know from \cite[Lemma 1]{FR02} that
\begin{equation}\label{EMa}
\chi(x^{(s+1)}>x^{(s)}) = \det [ \chi(x_j^{(s+1)} > x_k^{(s)}) ]_{j,k=1,\dots,s+1}
\end{equation}
where $x_{s+1}^{(s)} := - \infty$. Consequently (\ref{EM}) can be written in the form
\begin{equation}\label{EM1}
{1 \over {C}} \prod_{s=1}^{N-1} \det [ \phi(x_j^{(s)},x_k^{(s+1)})]_{j,k=1,\dots,s+1}
\det [ \Psi_{N-j}^N(x_k^{(N)})]_{j,k=1,\dots,N}
\end{equation}
with
\begin{equation}\label{EM2}
\phi(x,y) := \chi_{y > x}, \qquad
\Psi_{j}^N(x) := w(x) p_{j}(x).
\end{equation}

The general structure (\ref{EM1}) is precisely that for which the correlations have been determined
in the recent work \cite[Lemma 3.4]{BFPS06}. To apply this result, with $(a * b)(x,y) :=
\int_{-\infty}^\infty a(x,z) b(z,y) \, dz$, it is necessary to compute the quantities
$$
\phi^{(n_1,n_2)}(x,y) := \underbrace{( \phi * \cdots * \phi) }_{n_2 - n_1 \: {\rm times}} (x,y),
\qquad n_1 < n_2
$$
(for $n_1 \ge n_2$, $\phi^{(n_1,n_2)}(x,y) := 0$), and
$$
\Psi_{n-j}^n(x) := (\phi^{(n,N)} * \Psi_{N-j}^N)(x) \quad (1 \le n < N, \: j=1,\dots,N).
$$
Use of (\ref{EM2}) shows
\begin{equation}\label{EM3}
\phi^{(n_1,n_2)}(x,y) = {1 \over (n_2 - n_1 - 1)!} \chi_{y > x} (y - x)^{n_2 - n_1 - 1}
\end{equation}
(with the convention that $1/(-p)! = 0$ for $p \in \mathbb Z^+$, this vanishes for $n_1 \ge n_2$) and
\begin{equation}\label{3.1}
\Psi_{n-j}^n(x) = {1 \over (N-n-1)!} \int_x^\infty w(y) p_{N-j}(y) (y - x)^{N-n-1} \, dy.
\end{equation}

To proceed further, we choose $w(y)$ to be one of the classical weight functions
\begin{equation}\label{3.2}
w(y) = \left \{
\begin{array}{ll} e^{-y^2}, & {\rm Gaussian} \\
y^a e^{-y} \chi_{y>0}, & {\rm Laguerre} \\
y^a (1 - y)^b \chi_{0 < y < 1}, & {\rm Jacobi}. \end{array} \right.
\end{equation}
We further choose $p_j(y)$ to be proportional to the corresponding orthogonal polynomials, as
specified by their Rodrigues formulas
\begin{equation}\label{4.1}
p_j(y) = {1 \over e_j w(y) }
{d^j \over dy^j} \Big ( w(y) (Q(y))^j \Big ) =
\left \{ \begin{array}{ll} H_j(y), & {\rm Gaussian} \\
L_j^{(a)}(y), & {\rm Laguerre} \\
P_j^{(a,b)}(1-2y), & {\rm Jacobi} \end{array} \right.
\end{equation}
with the quantities $e_j$ and $Q(y)$ defined in the various cases by the pairs
\begin{equation}\label{4.1a}
(e_j, Q(y)) = 
\left \{ \begin{array}{ll} ((-1)^j, 1), & {\rm Gaussian} \\
(j!,y), & {\rm Laguerre} \\
(2^j j!,y(1-y)), & {\rm Jacobi}. \end{array} \right.
\end{equation}

Substituting (\ref{4.1}) in (\ref{3.1}) and integrating by parts shows that for $j \ge 0$
($n \ne N$)
\begin{equation}\label{4.2}
\Psi_j^n(x) = (-1)^{N-n} {e_j \over e_{N-n+j}}
\left \{ \begin{array}{ll} w(x) H_j(x), & {\rm Gaussian} \\
w(x) |_{a \mapsto a+N-n} L_j^{(a+N-n)}(x), & {\rm Laguerre} \\
w(x) |_{a \mapsto a+N-n \atop b \mapsto b+N-n} P_j^{(a+N-n,b+N-n)}(1-2x), & {\rm Jacobi}, \end{array} \right.
\end{equation}
while for $j<0$
\begin{equation}\label{5.0}
\Psi_j^{n}(x) = {(-1)^{N-n+j} \over e_{N-n+j} } {1 \over (-j-1)!}
\int_x^\infty ( y - x)^{-j-1} w(y) (Q(y))^{N-n+j} \, dy.
\end{equation}
As further required by \cite[Lemma 3.4]{BFPS06}, one introduces the polynomials 
$\{\Phi_j^n(x)\}_{j=0,\dots,n-1}$,
$n=1,\dots,N-1$
by the orthogonality requirement
$$
\int_{-\infty}^\infty \Phi_j^n(x) \Psi_k^n(x) \, dx = \delta_{j,k}.
$$
{}From (\ref{4.2}) we see that 
\begin{equation}\label{5.1}
\Phi_j^n(x) = (-1)^{N-n} {e_{N-n+j} \over e_{j}}
\left \{ \begin{array}{ll} {1 \over {\cal N}_j } H_j(x), & {\rm Gaussian} \\
{1 \over {\cal N}_j |_{a \mapsto a + N - n} } L_j^{(a+N-n)}(x), & {\rm Laguerre} \\
{1 \over {\cal N}_j |_{a \mapsto a+N-n \atop b \mapsto b+N-n} } 
P_j^{(a+N-n,b+N-n)}(1-2x), & {\rm Jacobi}, \end{array} \right.
\end{equation}
where
\begin{equation}\label{5.2}
{\cal N}_j =
\left \{ \begin{array}{ll} 2^j j! \sqrt{\pi},  & {\rm Gaussian} \\
{\Gamma(j+a+1) \over \Gamma(j+1)},  & {\rm Laguerre} \\
{ \Gamma(j+a+1) \Gamma(j+b+1) \over j! (2j+a+b+1) \Gamma(j+a+b+1) },
& {\rm Jacobi} \end{array} \right.
\end{equation}
is the normalization associated with each polynomial respectively.

A crucial feature exhibited by (\ref{5.1}) is that $\Phi_0^n(x)$ is a constant.
Now, Assumption (A) of  \cite[Lemma 3.4]{BFPS06} requires that $\Phi_0^n(x) \propto \phi(x_{n+1}^{(n)},x)$.
Recalling from below (\ref{EMa}) that $x_{n+1}^{(n)} := - \infty$, we see from the first definition
in (\ref{EM2}) that indeed $ \phi(x_{n+1}^{(n)},x)$ is similarly a constant. With 
Assumption (A) satisfied,  \cite[Eq.~(3.25)]{BFPS06} gives that the correlation between eigenvalues of
species $s_j$ at positions $y_j$  $(j=1,\dots,r)$ has the determinant form
\begin{equation}\label{6.1}
\rho(\{(s_j,y_j)\}_{j=1,\dots,r}) = \det [ K(s_j,y_j;s_k,y_k) ]_{j,k=1,\dots,r},
\end{equation}
with
the kernel $K$ given in terms of the quantities $\phi^{(n_1,n_2)}(x,y)$, $\Psi_j^n(x)$,
$\Phi_j^n(x)$ specified above according to
\begin{equation}\label{6.2}
K(s_j,y_j;s_k,y_k) = - \phi^{(s_j,s_k)}(y_j,y_k) +
\sum_{l=1}^{s_k} \Psi^{s_j}_{s_j - l}(y_j) 
\Phi_{s_k - l}^{s_k}(y_k).
\end{equation}

These findings can be summarized in the following statement.

\begin{prop}\label{p2}
Consider the joint PDF (\ref{EM}), with $w(y)$ one of the three classical weights (\ref{3.2}).
Specify $\phi^{(n_1,n_2)}(x,y)$ by (\ref{EM3});
$e_j, Q(y)$ by (\ref{4.1a}); $\Psi_j^n(x)$ by (\ref{4.2}), (\ref{5.0}); $\Phi_j^n(x)$ by (\ref{5.1}) and
${\cal N}_j$ by (\ref{5.2}). In terms of these quantities, the general $r$-point correlation is given
by (\ref{6.1}) with kernel (\ref{6.2}).
\end{prop}

\subsection{Direct approach}
Here the method of \cite{NF98} will be used to reclaim Proposition \ref{p2}. The starting
point is (\ref{EM1}), rewritten in the form
\begin{eqnarray}\label{j1.1}
&& {1 \over C} \prod_{s=2}^N
\det \left [ \begin{array}{cc} 1_{(N-s) \times (N-s)} & 0_{(N-s) \times s} \\
0_{s \times (N-s)} & \left [ \begin{array}{c}
[ \phi(x_j^{(s-1)}, x_k^{(s)}) - \kappa_s(x_j^{(s-1)}) ]_{j=1,\dots,s-1 \atop
k = 1,\dots,s}  \\{} [1]_{k=1,\dots,s} \end{array} \right ] \end{array} \right ]\nonumber \\
&& \qquad \times \det [ \psi_{j-1}^N(x_k^{(N)}) ]_{j,k=1,\dots,N}
\end{eqnarray}
so that all determinants are of the same dimension. The auxiliary function $\kappa_s(x)$
is arbitrary, the value of the determinant being independent of $\kappa_s(x)$.

In the notation for $w(y)$, $p_j(y)$ and ${\cal N}_j$ as defined by
(\ref{3.2}), (\ref{4.1}) and (\ref{5.2}), introduce the superscript $(N-n)$ to
indicate that $a \mapsto a + n$ (Laguerre case), $a \mapsto a + n$, $b \mapsto b + n$ 
(Jacobi case). In the Gaussian case the superscript has no effect. With this 
meaning understood,
expanding $\phi(x,y) = \chi_{y > x}$ in terms of $\{p_j^{(s)}(y)\}_{j=0,1,\dots}$ gives
\begin{equation}\label{pip}
\phi(x,y) = \sum_{k=0}^\infty {p_k^{(s)}(y) \over {\cal N}_k^{(s)} }
\int_x^\infty w^{(s)}(t) p_k^{(s)}(t) \, dt.
\end{equation}
Separating off the $k=0$ term, making use of the Rodrigues formula (\ref{4.1}) and
integrating by parts
reduces this to
\begin{equation}\label{pip1}
\phi(x,y) = {1 \over {\cal N}_0^{(s)} } \int_x^\infty w^{(s)}(t) \, dt -
 w^{(s-1)}(x) \sum_{k=0}^\infty {e_k \over e_{k+1} {\cal N}_{k+1}^{(s)} } p_k^{(s-1)}(x)
p_{k+1}^{(s)}(y).
\end{equation}
Recalling from (\ref{j1.1}) that $\kappa_s(x)$ is to be subtracted from
$\phi(x,y)$, the formula (\ref{pip1}) suggests choosing 
\begin{equation}\label{j1.2}
\kappa_s(x) = {1 \over {\cal N}_0^{(s)} } \int_x^\infty w^{(s)}(t) \, dt.
\end{equation}
Making this choice, and with the notation
\begin{equation}\label{ep}
\eta_k^{(s)}(x) = \bigg ( {w^{(s)}(x) \over {\cal N}_k^{(s)} } \bigg )^{1/2}
p_k^{(s)}(x)
\end{equation}
(note that $\{ \eta_k^{(s)}(x) \}_{k=0,1,\dots}$ is a set of orthonormal functions) we
obtain for (\ref{j1.1}) the expression
\begin{eqnarray}\label{j2.1}
&& {1 \over C} \prod_{s=2}^N
\det \left [ \begin{array}{cc} 1_{(N-s) \times (N-s)} & 0_{(N-s) \times s} \\
0_{s \times (N-s)} & \left [ \begin{array}{c}
{}[1]_{k=1,\dots,s} \\
{}[\tilde{\phi}_s(x_j^{(s-1)}, x_k^{(s)})]_{j=1,\dots,s-1 \atop
k = 1,\dots,s} \end{array} \right ] \end{array} \right ] \nonumber \\
&& \qquad \times \det [ \eta_{j-1}^{(N)}(x_k^{(N)}) ]_{j,k=1,\dots,N}
\end{eqnarray}
where, with 
\begin{equation}\label{gg}
\gamma_j^{(t)} := e_j ({\cal N}_j^{(t)} )^{1/2},
\end{equation}
the quantity $\tilde{\phi}_s$ is specified by
$$
\tilde{\phi}_s(x,y) := \sum_{k=0}^\infty {\gamma_k^{(s-1)} \over \gamma_{k+1}^{(s)} }
\eta_k^{(s-1)}(x) \eta_{k+1}^{(s)}(y)
$$
and for convenience the final row in the bottom right block of the first determinant
has been moved to the first row.

Our next step is to introduce
$$
\eta_{j,l}^{(s)} := \left \{ \begin{array}{ll}
\eta_j^{(s)}(x_l^{(s)}), & j \ge 0, \: l \ge 1 \\
\delta_{j,l-1}, & {\rm otherwise} \end{array} \right.
$$
and in terms of this to define
\begin{eqnarray*}
A_{j,l}^{(s,t)} & := & \sum_{k=0}^{N-1} {\gamma_{k+s-N}^{(s)} \over
\gamma_{k+t-N}^{(t)} } \eta_{k+s-N,j+s-N}^{(s)}  \eta_{k+t-N,l+t-N}^{(t)} \\
G_{j,l}^{(s,t)} & := & \sum_{k=0}^{\infty} {\gamma_{k+s-N}^{(s)} \over
\gamma_{k+t-N}^{(t)} } \eta_{k+s-N,j+s-N}^{(s)}  \eta_{k+t-N,l+t-N}^{(t)}.
\end{eqnarray*}
We can write (\ref{j2.1}) in terms of $\{A_{j,l}^{(s,t)} \}$, $\{ G_{j,l}^{(s,t)} \}$
so that it reads
\begin{eqnarray}\label{j3.1}
&& {1 \over C} \det [ A_{j,l}^{(N,1)} ]_{j,l=1,\dots,N}
\prod_{s=2}^N \det [ G_{j,l}^{(s-1,s)} ]_{j,l=1,\dots,N} \nonumber \\
&& \qquad =
{1 \over C} \det
\left [ \begin{array}{cccccc}
A^{(N,1)} & A^{(N,2)} & A^{(N,3)} & A^{(N,4)} & \cdots & A^{(N,N)} \\
0 & - G^{(1,2)} & -  G^{(1,3)} & - G^{(1,4)} & \cdots & - G^{(1,N)} \\
0 & 0 & - G^{(2,3)} & - G^{(2,4)} & \cdots & - G^{(2,N)} \\
0 & 0 & 0 & - G^{(3,4)} & \cdots & - G^{(3,N)} \\
\vdots & \vdots & \vdots & \vdots & \ddots & \vdots \\
0 & 0 & 0 & 0 & \cdots & - G^{(N-1,N)} \end{array} \right ]
\end{eqnarray}
where $A^{(s,t)} := [ A_{j,l}^{(s,t)} ]_{j,l=1,\dots,N}$,
$ G^{(s,t)} = [ G_{j,l}^{(s,t)} ]_{j,l=1,\dots,N}$. With $\alpha^{(s,t)}$ the $N \times N$
matrix such that $\alpha^{(s,t)} A^{(t,u)} = A^{(s,u)}$, multiply row 1 by
$\alpha^{(j-1,N)}$ and add to row $j$ $(j=2,\dots,N)$ to rewrite this as
$$
{1 \over C} \det \left [
\begin{array}{cccccc} A^{(N,1)} & A^{(N,2)} & A^{(N,3)} & A^{(N,4)} & \cdots & 
A^{(N,N)} \\
A^{(1,1)} & B^{(1,2)} & B^{(1,3)} & B^{(1,4)} & \cdots & B^{(1,N)} \\
A^{(2,1)} & A^{(2,2)} & B^{(2,3)} & B^{(2,4)} & \cdots & B^{(2,N)} \\
A^{(3,1)} & A^{(3,2)} & A^{(3,3)} & B^{(3,4)} & \cdots & B^{(3,N)} \\
\vdots & \vdots & \vdots & \vdots & \ddots & \vdots \\
A^{(N-1,1)} &  A^{(N-1,2)} & A^{(N-1,3)} &  A^{(N-1,4)} & \hdots &
B^{(N-1,N)} \end{array} \right ]
$$
where $B^{(s,t)} := A^{(s,t)} - G^{(s,t)}$. Moving the first block-row to the final
block-row gives the structured formula
\begin{equation}\label{18}
{1 \over C} \det [ F^{(s,t)} ]_{s,t=1,\dots,N}, \qquad
F^{(s,t)} := \left \{ \begin{array}{ll} A^{(s,t)}, & s \ge t \\
B^{(s,t)}, & s < t. \end{array} \right.
\end{equation}

Moreover, from the definition of $A^{(s,t)}$ and $G^{(s,t)}$ we observe that for
$s \ge t$
$$
F_{j,l}^{(s,t)} = \delta_{j,l}, \qquad j \le N - s \: \: {\rm or} \: \: l \le N -s
$$
while for $s < t$
$$
F_{j,l}^{(s,t)} = 0, \qquad j \le N - s \: \: {\rm or} \: \: l \le N - t.
$$
This allows the dimension of the block matrix in (\ref{18}) to be reduced, giving
for the joint PDF (\ref{EM}), $p$ say,
\begin{equation}\label{19}
p(\vec{x}^{(1)},\dots, \vec{x}^{(N)}) = {1 \over C} \det [ f^{(s,t)} ]_{s,t=1,\dots,N}
\end{equation}
where $\vec{x}^{(j)} = (x_1^{(j)}, \dots, x_j^{(j)})$ and
$f^{(s,t)}$ is the $s \times t$ matrix with entries
\begin{equation}\label{fF}
f_{j,l}^{(s,t)} = F_{j-s+N, l - t + N}^{(s,t)} =
\left \{
\begin{array}{ll}
\sum_{k=1}^t {\gamma^{(s)}_{s-k} \over \gamma^{(t)}_{t-k} } \eta_{s - k}^{(s)}(x_j^{(s)})
\eta_{t - k}^{(t)}(x_l^{(t)}), & s \ge t \\
- \sum_{k= - \infty}^0  {\gamma^{(s)}_{s-k} \over \gamma^{(t)}_{t-k} } \eta_{s - k}^{(s)}(x_j^{(s)})
\eta_{t - k}^{(t)}(x_l^{(t)}), & s < t. \end{array} \right.
\end{equation}

{}From the orthonormality of $\{\eta_k^{(s)}(x) \}$ it follows from
(\ref{19}) that
\begin{equation}\label{20a}
\int_{-\infty}^\infty f_{j,l}^{(s,t)} f_{l,m}^{(t,u)} \, dx_l^{(t)} =
\left \{ \begin{array}{ll} f_{j,m}^{(s,u)}, & s \ge t \ge u \: \: {\rm or}
\: \: s < t < u \\ 0, & {\rm otherwise}. \end{array} \right.
\end{equation}
We seek to use the form (\ref{19}), together with the property
(\ref{20a}), to compute the correlation between
eigenvalues of species $s_j$ at positions $y_j$ $(j=1,\dots,r)$. For this
purpose, we group together the eigenvalues of distinct species in the
correlation. Thus if the distinct species are $\hat{s}_1,\dots,\hat{s}_{\hat{r}}$,
with $\hat{s}_1,\dots, \hat{s}_{\hat{r}} \in \{ s_1,\dots, s_r \}$, we write
the positions being observed in species $\hat{s}$ as
$\vec{x}^{(\hat{s})} := ( x_1^{(\hat{s})}, \dots , x_{n_{\hat{s}}}^{(\hat{s})} )$
$(1 \le n_{\hat{s}} \le \hat{s})$. The correlation relating to
$\{ \vec{x}^{( \hat{s}_j )} \}_{j=1,\dots, \hat{r} }$ is specified in terms of 
the PDF $p$ by
\begin{eqnarray}\label{18b}
&& \rho( \{ \vec{x}^{(\hat{s}_j)} \}_{j=1,\dots, \hat{r} } ) =
\bigg ( \prod_{s = 1 \atop s \notin \{ \hat{s}_1, \dots, \hat{s}_{\hat{r}} \} }^N
\int d x_1^{(s)} \cdots \int d x_s^{(s)} \bigg ) \bigg (
\prod_{a=1}^{\hat{r} }  {\hat{s}_a! \over n_{\hat{s}_a}! } 
\int d x^{\hat{s}_a}_{n_{\hat{s}_a} + 1}  \cdots \int d x_{\hat{s}_a}^{(\hat{s}_a)} \bigg )
\nonumber \\
&& \qquad \times p(\vec{x}^{(1)},\dots, \vec{x}^{(N)} ).
\end{eqnarray}

Because of the structure (\ref{19}) and the orthogonality relation
(\ref{20a}), these integrals can all be computed by performing a Laplace
expansion of the determinant (see e.g.~\cite{NF98,Fo02}) to give
$$
 \rho( \{ \vec{x}^{(\hat{s}_j)} \}_{j=1,\dots, \hat{r} } ) =
\det \Big [ [f_{j, k }^{
(\hat{s}_\alpha, \hat{s}_\beta)} ]_{j=1,\dots,n_\alpha \atop k=1,\dots,n_\beta}
\Big ]_{\alpha, \beta = 1,\dots, \hat{r}}.
$$
In the notation of (\ref{6.1}) this can equivalently be written
\begin{equation}\label{4.28a}
\rho(\{(s_j,x_j)\}_{j=1,\dots,r}) = \det [ f_{j,k}^{(s_j, s_k)}
]_{j,k=1,\dots,r}
\end{equation}
(note that the superscript on $x_j^{(s_j)}$ is now redundant, and hence
has been omitted)
so to complete our task of rederiving (\ref{6.1}) it is sufficient to show
that
\begin{equation}\label{ffs}
 f_{j,k}^{s_j, s_k} = {a(s_j,x_j) \over a(s_k, x_k)} K(s_j, x_j; s_k, x_k)
\end{equation}
(the corresponding determinant is independent of the function $a(s,x)$).

To verify (\ref{ffs}) we begin by recalling (\ref{ep}) and (\ref{gg}) to see
from the definition (\ref{fF}) that for $s \ge t$
\begin{equation}\label{s1a}
f_{j,l}^{(s,t)} = \Big ( w^{(s)}(x_j) w^{(t)}(x_l) \Big )^{1/2}
\sum_{k=1}^t {e_{s-k} \over e_{t-k} }
{p_{s-k}^{(s)} (x_j) p_{t-k}^{(t)} (x_l) \over {\cal N}_{t-k}^{(t)} }
\end{equation}
and for $s < t$
\begin{equation}\label{s2a}
f_{j,l}^{(s,t)} = - \Big ( w^{(s)}(x_j) w^{(t)}(x_l) \Big )^{1/2}
\sum_{k=-\infty}^0 {e_{s-k} \over e_{t-k} }
{p_{s-k}^{(s)} (x_j) p_{t-k}^{(t)} (x_l) \over {\cal N}_{t-k}^{(t)} }.
\end{equation}
On the other hand, it follows from (\ref{4.1}) and (\ref{4.2}) that for $j \ge 0$
$(n \ne N)$
$$
\psi_j^n(x) = (-1)^{N-n} {e_j \over e_{N-n+j}} w^{(n)}(x) p_j^{(n)}(x),
$$
while according to (\ref{5.1})
\begin{equation}\label{sus}
\Phi_j^n(x) = (-1)^{N-n} {e_{N - n + j} \over e_j} {p_j^{(n)}(x) \over 
{\cal N}_j^{(n)} }.
\end{equation}
Recalling too that $\phi^{(s,t)}(x,y) = 0$ for $ s \ge t$ we then see from the
definition (\ref{6.2}) that for $s \ge t$
\begin{equation}\label{s1b}
K(s,x_j;t,x_l) = (-1)^{s-t} w^{(s)}(x_j)
\sum_{k=1}^t {e_{s-k} \over e_{t-k} }
{p_{s-k}^{(s)} (x_j) p_{t-k}^{(t)} (x_l) \over {\cal N}_{t-k}^{(t)} }.
\end{equation}

For $s < t$
$$
\phi^{(s,t)}(x_j, x_l) = {1 \over (t-s-1)!} \chi_{x_l > x_j}
(x_l - x_j)^{t-s-1}.
$$
Analogous to (\ref{pip}), we can expand $\phi^{(s,t)}$ in terms of $\{p_k^{(t)}(y) \}$,
$$
\phi^{(s,t)}(x,y) = {1 \over (t-s-1)!}
\sum_{k=0}^\infty {p_k^{(t)}(y) \over {\cal N}_k^{(t)}}
\int_x^\infty w^{(t)}(u) (u - x)^{t - s -1} p_k^{(t)}(u) \, du.
$$
Proceeding now as in the derivation of (\ref{pip1}) gives
\begin{eqnarray}\label{pop1}
&& \phi^{(s,t)}(x,y) = (-1)^{t - s } w^{(s)}(x) \sum_{k=-\infty}^s
{e_{s-k} \over {\cal N}_{t-k}^{(t)} e_{t-k} } p_{s-k}^{(s)}(x) p_{t-k}^{(t)}(y) \nonumber \\
&& \qquad + \sum_{k=0}^{t - s - 1} {(-1)^k \over (t - s - k - 1)! }
{p_k^{(t)}(y) \over e_k {\cal N}^{(t)}_k }
\int_x^\infty w^{(t-k)}(u) (u - x)^{t - s - 1 - k} \, du.
\end{eqnarray}
But from (\ref{5.0}), for $j < 0$
$$
\Psi_j^n(x) = {(-1)^{N-n+j} \over e_{N-n+j} } {1 \over (-j-1)!}
\int_x^\infty (u - x)^{-j-1} w^{(n-j)}(u) \, du
$$
so after making use too of (\ref{sus}) we have
\begin{eqnarray}\label{pop2}
&&
\sum_{p=s+1}^t \psi_{s-p}^s(x) \Phi_{t-p}^t(y) \nonumber \\
&& \qquad =
\sum_{k=0}^{t - s - 1} {(-1)^k \over (t - s - k - 1)!}
{p_k^{(t)}(y) \over e_k {\cal N}_k^{(t)} }
\int_x^\infty (u - x)^{t - s - 1 - k} w^{(t-k)}(u) \, dy.
\end{eqnarray} 
Thus adding (\ref{pop2}) to minus (\ref{pop1}) cancels the final line in the
latter. It remains to add to minus (\ref{pop1}) the quantity
\begin{equation}\label{4.35a}
\sum_{p=1}^s \psi_{s-p}^s(x) \Phi_{t-p}^t(y) = (-1)^{s-t}
w^s(x) \sum_{k=1}^s {e_{s-k} \over e_{t-k} }
{p_{s-k}^s(x) p_{t-k}^t(y) \over {\cal N}_{t-p}^{(t)} }.
\end{equation}
This cancels the corresponding terms in the first sum of minus (\ref{pop1}),
giving the result that for $s < t$
\begin{equation}\label{s2b}
K(s,x_j;t, x_l ) =  (-1)^{t - s - 1 } w^{(s)}(x) \sum_{k=-\infty}^0
{e_{s-k} \over {\cal N}_{t-k}^{(t)} e_{t-k} } p_{s-k}^{(s)}(x) p_{t-k}^{(t)}(y).
\end{equation}

Comparing (\ref{s1a}) with (\ref{s1b}), and (\ref{s2a}) with (\ref{s2b}), we see
that for general $s,t$
$$
f_{j,l}^{(s,t)} = (-1)^{s-t} \bigg ( {w^{(t)}(x_l) \over w^{(s)}(x_j) } \bigg )^{1/2}
K(s,x_j;t,x_l),
$$
thus verifying (\ref{ffs}). 

\section{Scaling limits}
\setcounter{equation}{0}
It is well known (see e.g.~\cite{Fo93a,Fo02}) that the eigenvalue distributions for the joint
PDF (\ref{CA1}) with the classical weights (\ref{3.2}) permit three distinct scalings as
$n \to \infty$. These correspond to eigenvalues in the bulk of the spectrum, or in the
neighbourhood of the spectrum edge. There are two distinct cases of the latter --- the soft edge
and the hard edge. The hard edge is characterized by the eigenvalue density being strictly zero
on one side. This occurs for $x < 0$ in the Laguerre ensemble, and for both $x<0$ and
$x > 1$ in the Jacobi ensemble. In contrast, the neighbourhood of the largest eigenvalue of the
Laguerre and Gaussian ensembles is such that the eigenvalue density is to leading order in $n$
zero, but at higher order it is non-zero. This is referred to as a soft edge.

For the projection process (\ref{EM}) with classical weights  (\ref{3.2}) we again
expect these same three distinct scalings, provided the difference between ranks of
the matrices (or equivalently between labels of the species) is fixed.
We find at the soft edge the correlations can be interpreted as
though the eigenvalues of the different species coincide with the eigenvalues of
species ($N$). Thus they are fully determined by the Airy kernel (see (\ref{KS})
below). That the eigenvalues of the different species should coincide at the soft
edge is not at all surprising upon consideration of  the joint PDF (\ref{KW}).
Thus one observes that the lowest indexed species repel via a Vandermonde factor,
with no restoring potential apart from the ordering constraint.
Thus they will tend to cluster at the boundaries, which at the soft edge corresponds to
the positions of the species $(N)$. 

The situation in the bulk and at the hard edge is more delicate in that the correlations depend on the
difference between labels of the species, even though this difference does not scale with
$N$. However the correlations within a given species can be anticipated. With
$w(x) = x^a e^{-x}$ in (\ref{EM}) the marginal distribution of species $N-c$ is precisely
the Laguerre unitary ensemble with $a \mapsto a+c$. Thus the hard edge correlations
within species $N-c$ must be the usual Bessel kernel correlations with $a \mapsto a+c$,
which is indeed what we find. Because the bulk correlations for the Laguerre and
Gaussian unitary ensembles are given by the sine kernel, all bulk correlations within
a species will be specified by the usual sine kernel. As to be discussed below,
for the bulk scaling it turns out that the full set of correlations are those known
from the so called bead process \cite{Bou06}.

Soft edge scaling, together with the difference between labels of the species scaling
with $N$ is of particular interest for its relevance to 
the queueing process of
Baryshnikov \cite{Ba01}, or equivalently a lattice version of the last passage
percolation model of Hammersley (see e.g.~\cite{Fo03}). To see this, let us revise some aspects
of the theory relating to the latter topic.

Thus, with each site
$(i,j)$ in the quadrant
$\mathbb Z^+ \times \mathbb Z^+$ an exponential random variable $x_{ij}$ of density
$2 e^{- 2 t}$, $t >0$. Define the stochastic variable
\begin{equation}\label{stv}
l(m,n) = {\rm max} \sum_{(1,1) \, {\rm u/r} \, (m,n) } x_{ij},
\end{equation}
where the sum is over all lattice paths in $\mathbb Z^+ \times \mathbb Z^+$ which start at
$(1,1)$ and finish at $(m,n)$ going either one lattice site up (u), or one lattice site to
the right (r). It is well known that with $x_n := l(n,n)$ the variables $\{x_1,\dots,x_n\}$ have
a joint PDF of the form (\ref{CA1}) with $w(x) = e^{-x}$ \cite{Jo99a}. Hence, as $n \to \infty$ the
corresponding distributions permit a soft edge scaling describing the scaled distribution of
$l(n,n)$. This scaling is fully described by the correlation function
\begin{equation}\label{Ak}
\rho^{\rm soft}_{(n)}(y_1,\dots,y_n) =
\det [ K^{\rm soft}(y_j,y_k) ]_{j,k=1,\dots,n}
\end{equation}
where 
\begin{eqnarray}\label{KS}
 K^{\rm soft}(x,y) & := & {{\rm Ai}(x) {\rm Ai}'(y) - {\rm Ai}(y) {\rm Ai}'(x) \over x - y } 
\nonumber \\
& = & \int_0^\infty {\rm Ai}(x + u) {\rm Ai}(y+u) \, du
\end{eqnarray}
is the so called Airy kernel.
It is also known \cite{Jo04}
that the sequence of stochastic variables $\{l(n+k,n-k) \}_{k=0,1,\dots}$ 
permit a scaling to a state specified by the dynamical extension of the correlation (\ref{Ak}),
\begin{equation}\label{Akd}
\rho^{\rm soft}_{(n)}((\tau_1,y_1),\dots,(\tau_n,y_n)) =
\det [ K^{\rm soft}((\tau_j,y_j),(\tau_k,y_k)) ]_{j,k=1,\dots,n}
\end{equation}
where
\begin{equation}\label{Kof}
K^{\rm soft}((\tau_x,x),(\tau_y,y)) = \left \{ \begin{array}{ll} A^{(1)}_{\tau_y - \tau_x}(x,y),
& \tau_y \ge \tau_x \nonumber \\ 
A^{(2)}_{\tau_y - \tau_x}(x,y), & \tau_y < \tau_x, \end{array} \right.
\end{equation}
\begin{eqnarray}\label{AA}
&& A^{(1)}_\tau(x,y) :=  \int_0^\infty e^{-\tau u}{\rm Ai}(x + u) {\rm Ai}(y+u)
\, du \nonumber \\
&& A^{(2)}_\tau(x,y) := - \int^0_{-\infty} e^{-\tau u}{\rm Ai}(x + u) {\rm Ai}(y+u)
\, du.
\end{eqnarray}
This is the so called Airy process ${\cal A}_2$, which underlies the distribution of the largest
eigenvalue in the scaled limit of the Dyson Brownian motion model of the GUE
\cite{Ma94,FNH98}.

The significance of these facts with respect to the present study is that in (\ref{KW}) with $n_2=0$,
$n_1 = p = n$, $w(y) = e^{-y}$, we know from the sentence including (\ref{2.19a}) that the variables
$\{y_{1}^{(n-p)}\}$ (i.e.~the largest eigenvalue for each species) coincide with the stochastic variables
$\{l(n,n-k)\}_{k=0,1,\dots}$. By analogy with the behaviour of the stochastic variables
$\{l(n+k,n-k) \}_{k=0,1,\dots}$, one may anticipate that their distribution is controlled
by the Airy process ${\cal A}_2$. We will find that with the differences between the ranks
(species) scaled to be of order $n^{2/3}$ that this is indeed the case, and that the same effect
holds for the soft edge in the Gaussian case.

\subsection{Fixed differences between species}
\subsection*{Soft edge and bulk scaling}
Explicit details will be worked out only in the
Gaussian case, as this case is typical; in particular the scaled
correlations do not depend on the particular case they originated 
from as is typical in random matrix theory (a form of universality).
In the $N \times N$ GUE the soft edge scaling is obtained
by the change of variables
\begin{equation}\label{xX}
x_i = \sqrt{2N} + {X_i \over \sqrt{2} N^{1/6} }.
\end{equation}
This has the effect of moving the origin to the neighbourhood of the largest eigenvalue,
and scaling the distances so the inter-eigenvalue spacings in this neighbourhood are
of order unity. The bulk scaling is obtained by the change of variables
\begin{equation}\label{uS}
x_i = {\pi X_i \over \sqrt{2N}}, 
\end{equation}
which makes the mean particle density in the neighbourhood of the origin unity. With the species
differing from $N$ by a constant,
\begin{equation}\label{ux}
s_i = N - c_i,
\end{equation}
we seek the limiting forms of the correlation
(\ref{6.1}) for both the soft edge and bulk scalings.

\begin{prop}\label{p1a}
For the soft edge scaling
\begin{equation}\label{xXa}
y_i = \sqrt{2N} + {Y_i \over \sqrt{2} N^{1/6} },
\end{equation}
and with $s_i$ specified in terms of $c_i$ by (\ref{ux}),
\begin{equation}\label{xXap}
{1  \over \sqrt{2} N^{1/6} } K(s_j,y_j;s_l,y_l ) 
\mathop{\sim}\limits_{N \to \infty}
{a_N(c_j, Y_j) \over a_N(c_l,Y_l) } K^{\rm soft}(Y_j, Y_l),
\end{equation}
where $K^{\rm soft}$ is given by (\ref{KS}) and $a_N(c,Y) :=
e^{-N^{1/3} Y} (2 N)^{-c/2}$. Consequently
\begin{equation}\label{5.11}
\lim_{N \to \infty} \Big ( {1 \over \sqrt{2} N^{1/6} } \Big )^r
\rho( \{ (s_j,y_j) \}_{j=1,\dots,r} ) =
\det [ K^{\rm soft}(Y_j,Y_k) ]_{j,k=1,\dots,r}.
\end{equation}

For the bulk scaling
\begin{equation}\label{us1}
y_i = {\pi Y_i \over \sqrt{2N} },
\end{equation}
and with $s_i$ specified in terms of $c_i$ by (\ref{ux}),
\begin{equation}\label{oy}
{\pi \over  \sqrt{2N} }
K(s_j,y_j;s_l,y_l ) 
\mathop{\sim}\limits_{N \to \infty} {b_N(c_j) \over b_N(c_l) }
B((c_j,Y_j),(c_l,Y_l)),
\end{equation}
where $b_N(c) :=  (2N)^{-c/2}$ and
\begin{equation}\label{5.13a}
B((\tau_x,x),(\tau_y,y)) := 
\left \{ \begin{array}{ll} \displaystyle
\int_0^1 s^{\tau_y - \tau_x} \cos ( \pi s (x - y) + \pi (\tau_x - \tau_y)/2) \, ds, &
 \tau_y \ge  \tau_x  \\ \displaystyle
-  \int_1^\infty s^{\tau_y - \tau_x} \cos ( \pi s (x - y) + \pi(\tau_x - \tau_y)/2) \, ds, &
 \tau_y < \tau_x \end{array}. \right.
\end{equation}
Consequently
\begin{equation}\label{5.14}
\lim_{N \to \infty} \Big ( {\pi \over \sqrt{2N}} \Big )^r
\rho( \{ (s_j,y_j) \}_{j=1,\dots,r} ) =
\det [ B((c_j,Y_j), (c_l,Y_l)) ]_{j,l=1,\dots,r}.
\end{equation}

\end{prop}

\noindent
Proof. \quad Substituting the Gaussian case of (\ref{3.2})--(\ref{4.1a}) and (\ref{5.2})
in (\ref{s1b}) shows that for $s_j \ge s_l$ ($c_j \le c_l$)
\begin{equation}\label{xX5a}
K(s_j, y_j; s_l, y_l ) = { e^{ - y_j^2 } \over \sqrt{\pi} }
\sum_{k=1}^{s_l} {1 \over 2^{s_l -k} (s_l - k) ! } H_{s_j - k}(y_j)
H_{s_l - k}(y_l ).
\end{equation}
Our strategy is to use appropriate expansions of the Hermite polynomials, corresponding
to the different scalings, to simplify the summation.

Consider first the soft edge scaling. With $x$ related to $X$ by (\ref{xX}) we have the
uniform large $N$ expansion \cite{Ol74}
\begin{equation}\label{xX5}
e^{ - x^2/2} H_N(x) = \pi^{1/4} 2^{N/2 + 1/4} (N!)^{1/2} N^{-1/12} \bigg (
{\rm Ai}(X) + {\rm O}(N^{-2/3}) \Big \{
\begin{array}{ll} {\rm O}(e^{-X}), & X > 0 \\
{\rm O}(1), & X < 0 \end{array} \bigg ).
\end{equation}
We rewrite this to read 
\begin{eqnarray}
&& e^{ - x^2/2} H_{N-k}(x) = \pi^{1/4} 2^{(N-k)2 + 1/4} ((N-k)!)^{1/2} N^{-1/12} 
\nonumber \\
&& \qquad \times
\bigg (
{\rm Ai}\Big (X + {k \over N^{1/3} } \Big ) + {\rm O}(N^{-2/3}) \Big \{
\begin{array}{ll} {\rm O}(e^{-k/N^{1/3}}), & k \ge 0 \\
{\rm O}(1), & k < 0 \end{array}  \bigg )
\end{eqnarray}
and then substitute in (\ref{xX5a}) to obtain
\begin{eqnarray}\label{xX6}
&&K(s_j, y_j; s_l, y_l ) \: \sim \:
e^{- N^{1/3} (Y_j - Y_k) } 2^{-(c_j - c_l)/2} 2^{1/2} s^{-1/6} \nonumber \\
&& \qquad \times
\sum_{k=1}^N \bigg ( {(N - c_j - k)! \over (N - c_l - k)! } \bigg )^{1/2}
{\rm Ai}(Y_j + k/N^{1/3}) {\rm Ai}(Y_l + k/N^{1/3}).
\end{eqnarray}

The leading order contribution to the sum in (\ref{xX6}) comes from terms 
${\rm o}(k^{1/3})$.
Noting that then $(N - c_j - k)!  / (N - c_l - k)! \sim N^{c_l - c_j}$ the sum can 
be recognised as the Riemann sum approximation to the integral form of $K^{\rm soft}$ in
(\ref{KS}), and (\ref{xXap}) in the case $c_j \le c_l$ follows.

We turn our attention next to analyzing (\ref{xX5a}) in the bulk scaling limit.
For this we use the uniform asymptotic expansion
$$
{\Gamma(n/2 + 1) \over \Gamma(n+1)} e^{-x^2/2} H_n(x) =
\cos ( \sqrt{2n+1} x - n \pi/2) + {\rm O}(n^{-1/2}),
$$
and a simple trigonometric identity to deduce that
\begin{eqnarray*}
&&K(s_j, y_j; s_l, y_l ) \: \sim \:
{1 \over 2 \sqrt{\pi} }
\sum_{k=1}^{N - c_l} {1 \over 2^{N - c_l - k} }
{(N - c_j - k)! \over ((N - c_j - k)/2)! ((N - c_l - k)/2)!} \nonumber \\
&& \qquad \times
\cos \Big ( \pi \sqrt{N-k \over N} (Y_j - Y_l) + {\pi \over 2} (c_j - c_l) \Big ).
\end{eqnarray*}
Here the main contribution to the sum comes from $(N-k)/N = {\rm O}(1)$. Expanding the ratio
of factorials in this setting gives 
\begin{eqnarray}\label{4.11}
&& K(s_j, y_j; s_l, y_l ) \: \sim \:
{2^{(c_l - c_j)/2} \over \sqrt{2} \pi} N^{(c_l - c_j - 1)/2}
\sum_{k=1}^{s_l} \Big ( {N - k \over N} \Big )^{(c_l - c_j - 1)/2} \nonumber \\
&& \qquad \times 
\cos \Big ( \pi \sqrt{N - k \over N} (Y_j - Y_l)  + {\pi \over 2} (c_j - c_l) \Big ).
\end{eqnarray}
Recognizing the sum as the Riemann sum approximation to a definite
integral in the variable $(N-k)/N = t$, then changing variables $t = s^2$ in the definite integral gives
(\ref{oy}) for $c_j \le c_l$. 

It remains to study the case $s_j < s_l$ ($c_j > c_l$)
for both the hard and soft scalings. For the
soft edge scaling it turns out that the form (\ref{s2b}) is not appropriate.
Instead we make use of the form (\ref{6.2}), which recalling (\ref{pop2}) and
(\ref{4.35a}) can be written
\begin{eqnarray}
&& K(s,x;t,y) = - \phi^{(s,t)}(x,y) + \sum_{k=0}^{t-s-1} {(-1)^k \over (t - s - k - 1)!}
{p_k^{(t)}(y) \over e_k {\cal N}_k^{(t)} } \int_x^\infty (u - x)^{t - s - 1 - k}
w^{(t-k)}(u) \, du \nonumber \\
&& \qquad + (-1)^{s-t} w^{(s)}(x) \sum_{k=1}^s {e_{s-k} \over e_{t-k} }
{p_{s-k}^{(s)}(x) p_{t-k}^{(t)}(y) \over {\cal N}_{t-k}^{(t)} }.
\end{eqnarray}
Considering this as the sum of three terms, the first two
do not contribute in
the scaling (\ref{xXa}), (\ref{ux}), and so
$$
K(s_j, y_j; s_l, y_l) \: \sim \:
(-1)^{s_j-s_l} w^{(s_j)}(y_j)
\sum_{k=1}^{s_j} {e_{s_j-k} \over e_{s_l-k} }
{p_{s_j-k}^{(s_j)} (y_j) p_{s_l-k}^{(s_l)} (y_l) \over {\cal N}_{s_l-k}^{(s_l)} }.
$$
This is precisely the expression (\ref{s1b}), except that the upper terminal is $s_j$
instead of $s_l$. Recalling the working below (\ref{xX5a}), this detail does not affect the
leading asymptotic form, so (\ref{xXap}) applies for both $c_j \le c_k$ and $c_j > c_k$.

In distinction to the strategy required at the soft edge for $s_j < s_l$
($c_j > c_l$) , to analyze the bulk
scaling in this case the form (\ref{s2b}) is well suited. As the only difference between
(\ref{s2b}) and (\ref{s1b}) is in the range of summation, the working leading to  
(\ref{4.11}) again applies, so this asymptotic formula remains valid but with
$k$ summed from $-\infty$ to 0. Crucially, because $c_j > c_l$ this sum is convergent
(it is in relation to this requirement that an analogous approach to the soft edge
scaling breaks down), and is furthermore a Riemann sum approximation to the same
definite integral as found for the case $c_j \le c_l$, but on $(-\infty,0]$
instead of $[0,1]$, hence implying the second formula in (\ref{oy}).
\hfill $\square$

\subsection{Hard edge scaling}
Hard edge scaling is possible for both the Laguerre and Jacobi cases;
here the details will be given in the Laguerre case only, as the limiting
correlations are the same in both cases. For the $N \times N$ LUE the hard
edge scaling results from the change of variables
\begin{equation}\label{ha}
x_i = {X_i \over 4N},
\end{equation}
which makes the inter-eigenvalue spacings in the neighbourhood of the hard edge
$x=0$ of order unity. We seek the limiting correlations with the scaling
(\ref{ha}) and the species specified by (\ref{ux}).

\begin{prop}
For the hard edge scaling (\ref{ha}) and with $s_i$ specified in terms of
$c_i$ by (\ref{ux})
\begin{equation}\label{hH}
{1 \over 4 N} K(s_j,x_j;s_l,x_l) \: \sim \:
{ h_N(c_j) \over h_N(c_l) } H(c_j,X_j;c_l,X_l),
\end{equation}
where $h_N(c) := (2N)^{-c}$ and
\begin{equation}\label{hH1}
H((\tau_x,x),(\tau_y,y)) :=
\left \{ \begin{array}{ll}\displaystyle
{1 \over 4} 
\int_0^1 s^{(\tau_y - \tau_x)/2} J_{a+\tau_x}((sx)^{1/2})
J_{a+\tau_y}((sy)^{1/2})\, ds, &
 \tau_y \ge  \tau_x  \\ \displaystyle
-  {1 \over 4} \int_1^\infty s^{(\tau_y - \tau_x)/2} J_{a+\tau_x}((sx)^{1/2})
J_{a+\tau_y}((sy)^{1/2}) \, ds, &
 \tau_y < \tau_x. \end{array} \right.
\end{equation}
Consequently
\begin{equation}
\lim_{N \to \infty} \Big ( {1 \over 4N} \Big )^r
\rho_{(r)}(\{(s_j,x_j)\}_{j=1,\dots,r}) = \det [ H(c_j,x_j;c_l,x_l)
]_{j,l=1,\dots,N}.
\end{equation}
\end{prop}

\noindent
{Proof. } \quad Substituting the explicit form of the Laguerre case of the
quantities in (\ref{s1b}) gives
\begin{equation}\label{sb1}
K(N- c_j,x_j; N - c_l, x_l ) =
x_j^{a + {c}_j} e^{- x_j}
\sum_{k=1}^{N- c_l} {\Gamma(N - {c}_j - k + 1) \over \Gamma(N - k +a + 1) }
L^{(a+{c}_j)}_{N - {c}_j - k} (x_j)
L^{(a+{c}_l)}_{N - {c}_l - k} (x_l),
\end{equation}
valid for $c_l \ge c_j$.
As $x_j, x_l$ are scaled according to (\ref{ha}), it is appropriate to make use
of the uniform asymptotic expansion \cite{Sz75}
$$
e^{-x/2} x^{a/2} L_n^a(x) = n^{a/2} J_a(2(nx)^{1/2}) + 
\bigg \{ \begin{array}{ll} x^{5/4} {\rm O}(n^{a/2 - 3/4}), & cn^{-1} < x < \omega \\
x^{a/2 + 2} {\rm O}(n^a), & 0 < x < c n^{-1}. \end{array} 
$$
Using this, and expanding the ratio of gamma functions with
$(N-k)/N = {\rm O}(1)$, we deduce
\begin{eqnarray}\label{fan}
&& K(N-c_j,x_j;N-c_l,x_l) \nonumber \\
&& \qquad \: \sim \: (2N)^{c_l - c_j}
\sum_{k=1}^{N-c_l} \bigg ( { N - k \over N} \bigg )^{(c_l - c_j)/2}
J_{a + c_j} \Big ( \Big ( {N - k \over N} X_j \Big )^{1/2} \Big )
J_{a + c_l} \Big ( \Big ( {N - k \over N} X_l \Big )^{1/2} \Big ). \nonumber \\
\end{eqnarray}
This is a Riemann sum, and the result (\ref{hH}) in the case
$c_l \ge c_j$ follows.

The expression (\ref{sb1}) is also valid for $c_l < c_j$, provided the summation
in now made over $k \in \mathbb Z_{\le 0}$. Following the above working through
again gives (\ref{fan}), but with the summation over $k \in \mathbb Z_{\le 0}$.
Because $c_l < c_j$ the sum is convergent, and its leading form given by
the definite integral made explicit in (\ref{hH1}).
\hfill $\square$

\subsection{Soft edge scaling with difference between species $O(N^{2/3})$}
As anticipated from the viewpoint of last passage percolation,
the soft edge scaling permits well defined correlations with the species separated
by O$(N^{2/3})$. The details can be worked out for both the Gaussian and
Laguerre cases, although the limiting correlations correspond to the Airy
process ${\cal A}_2$ and so are independent of the particular case.

\begin{prop}\label{p5}
In the Gaussian case, scale $s_i$ according to 
\begin{equation}
s_i = N + 2 c_i N^{2/3},
\end{equation}
and scale $y_i$ according to
\begin{equation}
y_i = (2 s_i)^{1/2} + {Y_i \over \sqrt{2} s_i^{1/6} }.
\end{equation}
For large $N$
\begin{equation}
{1 \over \sqrt{2} N^{1/6} } K(s_j,y_j; s_l, y_l) \: \sim \:
{\alpha_N(c_j,Y_j) \over \alpha_N(c_l,Y_l) }
\left \{ \begin{array}{ll} A^{(1)}_{c_j - c_l}(Y_j,Y_l), & c_j \ge c_l \\
 A^{(2)}_{c_j - c_l}(Y_j,Y_l), & c_j < c_l \end{array} \right.
\end{equation}
where $A^{(1)}, A^{(2)}$ are given by (\ref{AA}) and
$\alpha_N(c,Y) := e^{-N^{1/3} Y} (2N)^{c N^{2/3}} e^{N^{1/3} c^2} e^{-2 c^3/3}$.
Consequently
\begin{equation}
\lim_{N \to \infty}
\Big ( {1 \over \sqrt{2} N^{1/6} } \Big )^r \rho ( \{ (s_j,y_j)\}_{j=1,\dots,r}) =
\det [ K^{\rm soft}((-c_j,X_j),(-c_k,X_k)) ]_{j,k=1,\dots,r}.
\end{equation}

In the Laguerre case, scale $s_i$ according to
\begin{equation}\label{fr2}
s_i = N - \tilde{s}_i, \qquad \tilde{s}_i := 2 c_i (2N)^{2/3}
\end{equation}
and scale $y_i^{(s_i)}$ according to
\begin{equation}\label{fr1}
y_i^{(s_i)} = 4 s_i + 2(a+ N - s_i) + 2( 2N)^{1/3} Y_i.
\end{equation}
For large $N$
\begin{equation}\label{rr1}
2 (2 N)^{2/3}  K(s_j,y_j; s_l, y_l) \: \sim \:
{\beta_N(c_j,Y_j) \over \beta_N(c_l,Y_l)}
\left \{ \begin{array}{ll} A^{(1)}_{c_l - c_j}(Y_j,Y_l), & c_l \ge c_j \\
 A^{(2)}_{c_l - c_j}(Y_j,Y_l), & c_l < c_j \end{array} \right.
\end{equation}
with $\beta_N(c,Y) = e^{-(2N)^{1/3} Y} N^{- c (2N)^{2/3}} e^{2 (2N)^{1/3} c^2} e^{8 c^3/3}$. 
Consequently
\begin{equation}
\lim_{N \to \infty}
\Big ( 2 (2N)^{2/3} )  \Big )^r \rho ( \{ (s_j,y_j)\}_{j=1,\dots,r}) =
\det [ K^{\rm soft}((c_j,Y_j),(c_k,Y_k)) ]_{j,k=1,\dots,r}.
\end{equation}
\end{prop}

\noindent
Proof. \quad The derivation is very similar in both cases, so we'll be content
with presenting the details in the Laguerre case only. Reading off from
(\ref{sb1}) we have 
\begin{equation}\label{sb1a}
 K(N- \tilde{s}_j,y_j; N - \tilde{s}_l, y_l ) = 
y_j^{a + \tilde{s}_j} e^{- y_j}
\sum_{k=1}^{N- \tilde{s}_l} {\Gamma(N - \tilde{s}_j - k + 1) \over \Gamma(N - k +a + 1) }
L^{(a+\tilde{s}_j)}_{N - \tilde{s}_j - k} (y_j)
L^{(a+\tilde{s}_l)}_{N - \tilde{s}_l - k} (y_l).
\end{equation}
This formula is valid for $\tilde{s}_j \le \tilde{s}_l$ (for $\tilde{s}_j > \tilde{s}_l$ the
RHS is to be modified by multiplying by $-1$ and changing the summation terminals to
$k \in \mathbb Z_{\le 0}$; this modification does not change the working below in any
essential way, and so will not be considered explicitly).

We seek the asymptotic form of (\ref{sb1}) upon the scalings (\ref{fr1}) and (\ref{fr2}).
Adapting the strategy of the proof of Proposition
\ref{p1a}, our chief tool is the uniform asymptotic expansion \cite{Jo01}
\begin{eqnarray}\label{s15}
&&x^{a/2} e^{-x/2} L_n^a(x) = (-1)^n (2n)^{-1/3} \sqrt{(n+a)!/n!} \nonumber \\&&
\qquad \times \bigg (
{\rm Ai}(X) + {\rm O}(n^{-2/3}) \left \{
\begin{array}{ll} {\rm O}(e^{-X}), & X \ge 0 \\
{\rm O}(1), & X < 0 \end{array} \right. \bigg )
\end{eqnarray}
where 
\begin{equation}\label{anz}
x = 4n + 2a + 2(2n)^{1/3} X
\end{equation}
(this form allows for $a={\rm o}(n)$; the classical Plancherel-Rotach type formula
given in e.g.~\cite{Sz75} requires $a$ to be fixed and correspondingly has
$\sqrt{(n+a)!/n!}$ replaced by $n^{a/2}$). Use of this formula, rewritten to read
\begin{eqnarray}
&&x^{a/2} e^{-x/2} L_{n-k}^a(x) = (-1)^{n-k}
(2n)^{-1/3} \sqrt{(n-k+a)!/(n-k)!} \nonumber \\&&
\qquad \times \bigg (
{\rm Ai}\Big (X + {2k \over (2n)^{1/3} } \Big ) + {\rm O}(N^{-2/3}) \left \{
\begin{array}{ll} {\rm O}(e^{-k/n^{1/3}}), & k \ge 0 \\
{\rm O}(1), & k < 0 \end{array} \right. \bigg )
\end{eqnarray}
with $n = N - \tilde{s}_i$ shows that for large $N$
\begin{eqnarray*}
 &&K(N- \tilde{s}_j,y_j; N - \tilde{s}_l, y_l ) \sim
e^{- (2 N)^{1/3} (Y_j - Y_l) } 
\nonumber \\
&& \qquad \times 
(2N)^{-2/3}
\sum_{k=1}^{N - \tilde{s}_l} \bigg ( {(N - \tilde{s}_j - k)! \over (N - \tilde{s}_l - k)! }
\bigg )^{1/2} {\rm Ai}(Y_j + 2k/(2N)^{1/3}) {\rm Ai}(Y_l + 2k/(2N)^{1/3})
\end{eqnarray*}
(cf.~(\ref{xX6})). The leading order contribution to the summation comes from
$k$ of order $N^{1/3}$. Using this fact, 
noting from Stirling's formula that for large $s$
\begin{equation}\label{xX7}
\bigg ( {(s - k_j)! \over (s - k_l)!} \bigg )^{1/2} \sim
s^{(k_l - k_j)/2} e^{(k_j^2 - k_l^2)/4s} e^{(k_j^3 - k_l^3)/12 s^2},
\end{equation}
and using this formula with $s=N$, $k_i = k + \tilde{s}_i$ $(i=j,l)$
the sum is recognised as the Riemann sum approximation to $A^{(1)}$
as defined in (\ref{AA}), implying the result (\ref{rr1}) in the
case $ c_l \ge c_j$.
\hfill $\square$

\section{Discussion}
\setcounter{equation}{0}
As pointed out to us by A.~Borodin, replacing (\ref{5.13a}) by
\begin{equation}\label{5.14a}
\left \{
\begin{array}{ll} \displaystyle {1 \over 2 } \int_{-1}^1
(is)^{\tau_y - \tau_x} e^{i s \pi (x - y)} \, ds, & \tau_y > \tau_x \\
\displaystyle - {1 \over 2 } \int_{\mathbb R\backslash [-1,1]}
(is)^{\tau_y - \tau_x} e^{i s \pi (x - y)} \, ds, & \tau_y < \tau_x
\end{array} \right.
\end{equation}
leaves the determinant (\ref{5.14}) unchanged.
Further changing scale $x \mapsto x/\pi, y \mapsto y/\pi$ removes the
$\pi$'s from the exponents, and multiplies each integrand by a factor
of $1/\pi$ to account for the corresponding scaling of the correlation function.
The significance of this form is that it is identical to the $\gamma = 0$
case of the correlation kernel for the so-called bead model
\cite[Thm.~2]{Bou06}. In fact the bead model was already known to
be closely related to the GUE minor process \cite[Section 4.1]{Bou06}.
The form (\ref{5.14a}) can also be obtained as a limit of the incomplete
beta kernel of Okounkov and Reshetikhin \cite[Section 3.1.7]{OR01}
(write the parameter $z$ as $z = 1 + ia$, change variables $w=1+ias$, rescale
the space variable $l$ by $a^{-1}$ and take $a \to 0$). 
The recent work \cite{Go07} obtains the incomplete beta kernel in the context
of a study of random lozenge-tilings.
Further the kernels
of \cite[Thm.~4.4]{Bo06a} permit degeneracies to (\ref{5.14a}).

Another discussion point is in relation to consistency between the present
results, and results from \cite{FR02}.
In \cite{FR02} the correlations for the p.d.f.
\begin{equation}\label{psy}
{1 \over C} \prod_{j=1}^n e^{-(x_j + y_j)/2} e^{A(x_y - y_j)/2}
\prod_{1 \le j < k \le n} (x_j - x_k) (y_j - y_k)
\chi_{x_1 > y_1 > \cdots > x_n > y_n}
\end{equation}
were computed, along with the scaled limits at the soft and hard
edges, and in the bulk. The p.d.f.~(\ref{psy}) with 
$A = -1$, is identical to the p.d.f.~(\ref{KW}) with $w(y) = e^{-y}$,
$n_2 = n$, $p=1$ and
$y_{n+1}^{(1)} =0$. Setting $y_{n+1}^{(1)} = 0$ would not be expected to alter the soft edge
and bulk scaling limits, so it should be that the scaled correlations in
\cite{FR02} contain as special cases the results (\ref{5.11}) and (\ref{5.14})
for $|c_j - c_l | = 0,1$.

To see that this is indeed the case, we recall from \cite{FR02} that
with
\begin{equation}\label{psy1}
A \mapsto \left \{ \begin{array}{ll} \sqrt{n} \alpha/ \pi, & {\rm bulk} \\
\alpha/ 2 (2n)^{1/3}, & {\rm soft \: edge}, \end{array} \right.
\end{equation}
the scaled correlation for $k_1$ variables of species type $x$, and
$k_2$ variables of species type $y$ was calculated to equal
\begin{eqnarray}\label{psy1a}
&&
\rho_{(k_1,k_2)}(X_1,\dots,X_{k_1}; Y_1,\dots,Y_{k_2}) \nonumber \\
&& \qquad =
\det
\left [
\begin{array}{cc} [K_{\rm oo}^{\rm scaled}(X_j,X_l)]_{j,l=1,\dots,k_1}
&
 [K_{\rm oe}^{\rm scaled}(X_j,Y_l)]_{j=1,\dots,k_1 \atop l=1,\dots,k_2} \\
{}[K_{\rm eo}^{\rm scaled}(Y_j,X_l)]_{j=1,\dots,k_2 \atop l=1,\dots,k_1}
&
 [K_{\rm ee}^{\rm scaled}(Y_j,Y_l)]_{j,l=1,\dots,k_2 }
\end{array} \right ]
\end{eqnarray}
where
\begin{eqnarray}\label{psy2}
&& K_{\rm ee}^{\rm scaled}(Y,Y') = K^{\rm scaled}(Y,Y') \nonumber \\
&&  K_{\rm eo}^{\rm scaled}(Y,X) = - e^{\alpha(X-Y)} \chi_{X > Y} +
e^{\alpha X/2} \int_{-\infty}^X e^{- \alpha v/2} K^{\rm scaled}(v,Y) \, dv 
\nonumber \\
&&  K_{\rm oe}^{\rm scaled}(X,Y) = - e^{-\alpha X/2} {\partial \over \partial X}
\Big ( e^{\alpha X/2} K^{\rm scaled}(X,Y) \Big ) \nonumber \\
&& K_{\rm oo}^{\rm scaled}(X,X') = - e^{\alpha (X - X')/2}
{\partial \over \partial X} \Big ( e^{\alpha X/2} \int_{-\infty}^{X'}
e^{- \alpha v/2} K^{\rm scaled}(X,v) \, dv \Big ).
\end{eqnarray}
In the soft edge case $K^{\rm scaled} = K^{\rm soft}$ as specified by 
(\ref{KS}), while in the bulk $K^{\rm scaled} = K^{\rm bulk}$ where
\begin{equation}\label{psy3}
K^{\rm bulk}(X,Y) = {\sin \pi (X - Y) \over \pi (X - Y) }
= \int_0^1 \cos \pi (X - Y)t \, dt
\end{equation}

We see from (\ref{psy1}) that $A = -1$ corresponds to $\alpha = 0$ in the bulk, and
$\alpha \to - \infty$ at the soft edge. We see from (\ref{psy2}) that for
$\alpha \to -\infty$
\begin{eqnarray*}
&& K_{\rm eo}^{\rm soft}(Y,X) \: \sim \: - {2 \over \alpha}
K^{\rm soft}(Y,X), \qquad
 K_{\rm oe}^{\rm soft}(X,Y) \: \sim \: - {\alpha \over 2}K^{\rm soft}(X,Y), \nonumber \\
&& K_{\rm oo}^{\rm soft}(X,X') \: \sim \: K^{\rm soft}(X,X').
\end{eqnarray*}
When substituted in (\ref{psy1a}) the factors $-2/\alpha$, $-\alpha/2$ cancel,
and so agreement with (\ref{5.11}) is found. Further, setting $\alpha = 0$
in (\ref{psy2}) and recalling (\ref{psy3}) gives 
\begin{eqnarray*}
&& K_{\rm eo}^{\rm bulk}(Y,X) \Big |_{\alpha = 0} =
- \chi_{X > Y} + \int_{-\infty}^X {\sin \pi (v - Y) \over
\pi (v - Y) } \, dv, \\
&& K_{\rm oe}^{\rm bulk}(X,Y) \Big |_{\alpha = 0} =
\pi \int_0^1 t \sin \pi (X - Y)t \, dt, \qquad
K_{\rm oo}^{\rm scaled}(X,X') \Big |_{\alpha = 0} =
{\sin \pi (X - X') \over \pi (X - X')}.
\end{eqnarray*}
A simple calculation shows that the first of these can be rewritten
$$
K_{\rm eo}^{\rm bulk}(Y,X) \Big |_{\alpha = 0} = - \int_1^\infty {\sin \pi v (X
- Y) \over \pi v} \, dv.
$$
With this we obtain agreement with (\ref{5.14}) in the case $|c_j - c_l| = 0,1$, as
expected.

\subsection*{Acknowledgements}
We thank Eric Nordenstam for some stimulating discussions, and Alexei Borodin for
the content of the first paragraph on Section 6.
The work of PJF is supported by the Australian Research Council.

\section*{Appendix}
\renewcommand{\theequation}{A.\arabic{equation}}
\setcounter{equation}{0}
Since the completion of this work, Borodin and P\'ech\'e have posted a work
\cite{BP07} on the arXiv which, amongst other results, establishes our
Proposition \ref{p5}. The strategy used is, at a technical level, quite different
to that adopted here.

In this appendix we concern ourselves with another aspect of the work \cite{BP07},
relating 
to a generalization of our (\ref{A1}),
\begin{equation}\label{Ax}
A_{(n+1)} = A_{(n)} + \vec{x}_{(n)}  \vec{x}_{(n)}^\dagger, \qquad
A_{(0)} = [0]_{p \times p}
\end{equation}
where $\vec{x}_{(n)}$ is a $p \times 1$ column vector of complex Gaussians
with entries such that the modulus of the $i$-th component has distribution
$\Gamma[1,1/(\pi_{i} + \hat{\pi}_n)]$. As in Section \ref{s3.2},
the point of interest is in the joint eigenvalue PDF for $\{A_1,\dots,A_p\}$.
This is not computed directly, but as in the discussion around 
(\ref{2.19a}), it is noted that the directed percolation in the $p \times p$
square which each lattice site $(i,j)$ containing an exponential random
variable of density $(\pi_i + \hat{\pi}_j) e^{-(\pi_i + \hat{\pi}_j)s}$ has the
distribution of the stochastic variable $l(p,p)$ equal to that of the
distribution of the largest eigenvalue of
$A_{(p)}$. For the percolation problem, the joint distribution of
$\{l(j,p)\}_{j=1,\dots,p}$ can be calculated, leading to a 
joint PDF for the $p$ species of variables 
$\{x_j^{(s)}\}$, $(s=1,\dots,p)$ with $j=1,\dots,s$
proportional to
\begin{equation}\label{f1f}
\det [ e^{- \pi_i x_j^{(p)} }]_{i,j=1,\dots,p}
\prod_{k=1}^{p-1} \det \Big [ e^{- \hat{\pi}_{k+1} (x_j^{(k+1)} - x_i^{(k)})}
\chi_{x_j^{(k+1)} > x_i^{(k)} }\Big ]_{i,j=1,\dots,k+1} e^{- \hat{\pi}_1 x_1^{(1)}}.
\end{equation}
The question is asked as to whether this joint PDF is in fact the joint PDF
for the eigenvalues of the matrices $A_{(s)}$, $s=1,\dots,p$.

In fact the working from \cite[Section 5]{FR02b} allows this question to be
answered in the affirmative in the limit $\pi_i \to c $, $(i=1,\dots,p)$.
Thus it follows from  \cite[Corollary 3]{FR02b} that the condition PDF for the
eigenvalues $\{a_j\}_{j=1,\dots,n}$ of $A_{(n)}$ is proportional to
\begin{equation}\label{ff}
\prod_{i=1}^{n+1} \lambda_i^{p-(n+1)} \prod_{j=1}^n {1 \over a_j^{p-n} }
e^{- (c + \hat{\pi}_n) (\sum_{j=1}^{n+1} \lambda_j - \sum_{j=1}^n a_j) }
{\prod_{i<j}^{n+1} (\lambda_j - \lambda_i) \over \prod_{i<j}^n (a_j - a_i) }
\chi(\lambda < a).
\end{equation}
Let us now write $\lambda_j \mapsto \lambda_j^{(n+1)}$, $a_j \mapsto \lambda_j^{(n)}$.
The sort joint PDF is the product from $n=1,\dots,p-1$ of the conditional PDFs
(\ref{ff}), multiplied by the PDF in the case $n=1$, which is proportional to
$(x_1^{(1)})^{p-1} e^{- (c + \hat{\pi}_1) x_1^{(1)}}$. This gives (\ref{f1f}) with the
first determinant therein replaced by $\prod_{l=1}^p e^{- c x_l}
\prod_{i<j}^p(x_j^{(p)} - x_i^{(p)})$, thus verifying (\ref{f1f}) in the
case that $\pi_i \to c$, $(i=1,\dots,p)$.


\providecommand{\bysame}{\leavevmode\hbox to3em{\hrulefill}\thinspace}
\providecommand{\MR}{\relax\ifhmode\unskip\space\fi MR }
\providecommand{\MRhref}[2]{%
  \href{http://www.ams.org/mathscinet-getitem?mr=#1}{#2}
}
\providecommand{\href}[2]{#2}

\end{document}